\documentclass[aps,prb,twocolumn,letterpaper]{revtex4-1}
\usepackage[utf8]{inputenc}
\usepackage{graphicx,amsmath}
\usepackage{amssymb}
\usepackage{amsthm}

\begin{document}
\title{From the adiabatic theorem of quantum mechanics to topological states of matter}
\author{Jan Carl Budich$^{1,2}$ and Bj\"orn Trauzettel$^2$}

 \affiliation{$^1$Department of Physics, Stockholm University, Se-106 91 Stockholm, Sweden;\\
 $^2$Institute for Theoretical Physics and Astrophysics,
 University of W$\ddot{u}$rzburg, 97074 W$\ddot{u}$rzburg, Germany}
\date{\today}

\begin{abstract}Owing to the enormous interest the rapidly growing field of topological states of matter (TSM) has attracted in recent years, the main focus of this review is on the theoretical foundations of TSM. Starting from the adiabatic theorem of quantum mechanics which we present from a geometrical perspective, the concept of TSM is introduced to distinguish gapped many body ground states that have representatives within the class of non-interacting systems and mean field superconductors, respectively, regarding their global geometrical features. These classifying features are topological invariants defined in terms of the adiabatic curvature of these bulk insulating systems. We review the general classification of TSM in all symmetry classes in the framework of K-Theory. Furthermore, we outline how interactions and disorder can 
be included into the theoretical framework of TSM by reformulating the relevant topological invariants in terms of the single particle Green's function and by introducing twisted boundary conditions, respectively. We finally integrate the field of TSM into a broader context by distinguishing TSM from the concept of topological order which has been introduced to study fractional quantum Hall systems.
\end{abstract}
\maketitle

\section{Introduction}
Stimulated by the theoretical prediction \cite{KaneMele2005a,KaneMele2005b,BHZ2006} and experimental discovery \cite{koenig2007} of the quantum spin Hall (QSH) state, tremendous interest has been recently attracted by the study of topologocical properties of non-interacting band structures \cite{KaneHasan,RyuLudwig,XLReview2010}. A topological state of matter (TSM) can be understood as a non-interacting band insulator which is topologically distinct from a conventional insulator. This means that a TSM cannot be adiabatically, i.e., while maintaining a finite bulk gap, deformed into a conventional insulator without breaking its fundamental symmetries. This classification approach provides a new paradigm in condensed matter physics which goes beyond the mechanism of local order parameters associated with spontaneous symmetry breaking \cite{AndersonBasic}:~Different TSM differ in the value of their defining topological invariant but concur in all conventional symmetries. Interestingly, these global topological features are not 
always immediately visible in the microscopic equations of motion. However, the bulk topology leads to unique finite size effects at the boundary of a finite sample which has been coined bulk boundary correspondence \cite{Halperin1982,Volovik,XLReview2010}. This general mechanism gives rise to peculiar holographic transport properties of topologically non-trivial systems. Predicting and probing the rich phenomenology of these topological boundary effects in mesoscopic samples has become one of the most rapidly growing fields in condensed matter physics in recent years.\\

The focus of this review is threefold. First, we view the topological invariants characterizing TSM as the global analogues of geometric phases associated with the adiabatic evolution of a physical system \cite{Kato1950,Berry,WilczekZee1984}. Geometric phases are well known to reflect the local inner-geometric properties of the Hilbert space of a system with a parameter dependent Hamiltonian \cite{Simon1983,BohmBerryBook}. The topological invariants represent global features of a Bloch Hamiltonian, i.e., of a system whose parameter space is its $k$-space. This approach integrates TSM and physical phenomena related to geometric phases into a common theoretical context. Second, we present the topological classification of TSM in the framework of K-Theory \cite{Karoubi} in a more accessible way than in the pioneering work by Kitaev \cite{KitaevPeriodic}. Third, we discuss very recent developments concerning the generalization of the relevant topological invariants to interacting and disordered systems several 
of which were published after existing reviews on TSM. 
\section{The adiabatic theorem}
\label{sec:adtheo}
The gist of the adiabatic theorem can be understood at a very intuitive level: Once prepared in an instantaneous eigenstate with an eigenvalue which is separated from the neighboring states by a finite energy gap $\Delta$, the system can only leave this state via a transition which costs a finite excitation energy $\Delta$. A simple way to estimate whether such a transition is possible is to look at the Fourier transform $\tilde{\mathcal H}(\omega)$~of the time dependent Hamiltonian $\mathcal H(t)$. If the time dependence of $\mathcal H$~is made sufficiently slow, $\tilde{\mathcal H}(\omega)$~will only have finite matrix elements for $\omega\ll \Delta$. In this regime the system will stick to the same instantaneous eigenstate. This behavior is known as the adiabatic assumption.

\subsection{Proof due to Born and Fock}
The latter rather intuitive argument is at the heart of the adiabatic theorem of quantum mechanics which has been first proven by Born and Fock in 1928 \cite{Born1928} for non-degenerate systems. Let $\left\{\lvert n(t)\rangle\right\}_n$~be an orthonormal set of instantaneous eigenstates of $\mathcal H(t)$~with eigenvalues $\left\{E_n(t)\right\}_n$. The exact solution of the Schr\"odinger equation can be generally expressed as
\begin{align}
\lvert \Psi(t)\rangle = \sum_n c_n(t)\lvert n(t)\rangle \text{e}^{-i\phi^n_D(t)},
\label{eqn:bornansatz}
\end{align}
where the dynamical phase $\phi_D^n(t)=\int_{t_0}^tE_n(\tau)\text{d}\tau$~has been separated from the coefficients $c_n(t)$~for later convenience. Plugging Eq. (\ref{eqn:bornansatz}) into the Schr\"odinger equation yields
\begin{align}
\dot c_n = -c_n \langle n\rvert \frac{d}{dt} \lvert n\rangle - \sum_{m\ne n} c_m \frac{\langle n \rvert \left(\frac{d}{dt} \mathcal H\right)\vert m\rangle}{E_m-E_n}\text{e}^{i\phi^{nm}_D(t)}
\label{eqn:bornresult}
\end{align}
with $\phi^{nm}_D(t)=\phi^n_D(t)-\phi^m_D(t)$. The salient consequence of the adiabatic theorem is that the last term in Eq. (\ref{eqn:bornresult}) can be neglected in the adiabatic limit since its denominator $\lvert E_n-E_m\rvert\ge \Delta$~is finite whereas the matrix elements of $\frac{d}{dt} \mathcal H$~become arbitrarily small. More precisely, if we represent the physical time as $t=Ts$, where $s$~is of order $1$~for a change in the Hamiltonian of order $\Delta$~and $T$~is the large adiabatic timescale, then $\frac{d}{dt}=\frac{1}{T}\frac{d}{ds}$. Now, $\frac{d}{ds}\mathcal H\left(t(s)\right)$~is by construction of order $\Delta$. The entire last term in Eq. (\ref{eqn:bornresult}) is thus of order $\frac{1}{T}$. Under these conditions \footnote{As a minor technical point, we note that the proof by Born and Fock \cite{Born1928} also takes into account level crossings at isolated points. These slightly more general conditions are not of relevance for our purposes as we will only discuss fully gapped systems. More recent work by Avron and coworkers \cite{AvronGapless} reported a proof of the adiabatic theorem which, under certain conditions on the level spectrum, works without any gap condition.}, Born and Fock \cite{Born1928}~showed that the contribution of this second term vanishes in the adiabatic limit $T\rightarrow \infty$. Note that this is not a trivial result since the differential equation (\ref{eqn:bornresult}) is supposed to be integrated from $t=0$~to $t\sim T$, so that one could naively expect a contribution of order $1$~from a coefficient that scales like $1/T$.
The coefficient of $c_n$~in the first term on the right hand side of Eq. (\ref{eqn:bornresult}) is purely imaginary since $0= \frac{d}{dt}\langle n \vert n\rangle= (\frac{d}{dt}\langle n\rvert)\lvert n\rangle + \langle n\rvert \frac{d}{dt}\lvert n\rangle$~and hence does not change the modulus of $c_n$~when the differential equation $\dot c_n = -c_n \langle n\rvert\frac{d}{dt} \lvert n\rangle$~is solved as
\begin{align}
c_n(t)= c_n(t_0)~\text{e}^{-\int_{t_0}^t\langle n\rvert \frac{d}{d\tau} \lvert n\rangle \text{d}\tau}.
\label{eqn:bornberry}
\end{align}
Born and Fock \cite{Born1928} argue that $\langle n\rvert \frac{d}{dt} \lvert n \rangle=0~\forall_t$~amounts to a choice of phase for the eigenstates and therefore neglect also the first term on the right hand side of Eq. (\ref{eqn:bornresult}).\\

This review article is mainly concerned with physical phenomena associated with corrections to this in general unjustified assumption.  

\subsection{Notion of the geometric phase}

By neglecting the first term on the right hand side of Eq. (\ref{eqn:bornresult}), Ref. \onlinecite{Born1928} overlooks the potentially nontrivial adiabatic evolution, known as Berry's phase \cite{Berry}, associated with a cyclic time dependence of $\mathcal H$.
After a period $\left[0,T\right]$~of such a cyclic evolution, Eq. (\ref{eqn:bornberry}) yields
\begin{align}
c_n(T)= c_n(0)\text{e}^{-\oint_{0}^T\langle n\rvert \frac{d}{d\tau} \lvert n\rangle \text{d}\tau}.
\label{eqn:berryhidden}
\end{align}
To understand why the phase factor $\text{e}^{-\oint_{0}^T\langle n\rvert \frac{d}{d\tau} \lvert n\rangle \text{d}\tau}$~can in general not be gauged away, we remember that the Hamiltonian depends on time via the time dependence $R(t)$~of some external control parameters. Hence, $\langle n\rvert\frac{d}{dt} \lvert n\rangle= \langle n\rvert \partial_\mu \lvert n\rangle \dot R^\mu$, where $\partial_\mu = \frac{\partial}{\partial R^\mu}$. To reveal the mathematical structure of the latter expression, we define
\begin{align}
\mathcal A^{B}\left(\frac{d}{dt}\right)= \mathcal A^B_\mu \dot R^\mu = -i\langle n\rvert \partial_\mu \lvert n\rangle \dot R^\mu,
\label{eqn:Berrycon}
\end{align}
where $\mathcal A^B= A^B_\mu dR^\mu$~is called Berry's connection. $\mathcal A^B$~clearly has the structure of a gauge field: Under the local gauge transformation $\lvert n\rangle \rightarrow \text{e}^{i\xi}\lvert n\rangle$~with a smooth function $R\mapsto \xi(R)$, Berry's connection transforms like
\begin{align*}
\mathcal A^{B} \rightarrow \mathcal A^{B}+d\xi.
\end{align*}
Furthermore, the cyclic evolution defines a loop $\gamma:t\mapsto R(t),~t\in\left[0,T\right],~R(0)=R(T)$~in the parameter manifold $\mathcal R$. If $\gamma$~ can be expressed as the boundary of some piece of surface, then, using the theorem of Stokes, we can calculate
\begin{align}
-i\oint_{0}^T\langle n\rvert \frac{d}{d\tau} \lvert n\rangle \text{d}\tau=\int_\gamma \mathcal A^B = \int_{\mathcal S} d \mathcal A^B = \int_{\mathcal S} \mathcal F^B,
\label{eqn:StokesBerry}
\end{align}
where in the last step Berry's curvature $\mathcal F^B= \mathcal F^B_{\mu\nu}dR^\mu \wedge dR^\nu$~is defined as
\begin{align*}
\mathcal F_{\mu\nu}^B= -i\left(\langle \partial_\mu n\rvert \partial_\nu n\rangle- \langle \partial_\nu n\rvert \partial_\mu n\rangle\right)= 2\text{Im}\left\{\langle \partial_\mu n\rvert \partial_\nu n\rangle\right\}
\end{align*}
with the shorthand notation $\lvert \partial_\mu n\rangle = \partial_\mu \lvert n\rangle$. Note that $\mathcal F^B$~is a gauge invariant quantity that is analogous to the field strength tensor in electrodynamics. Defining the Berry phase associated with the loop $\gamma$~as $\varphi^B_\gamma= \int_\gamma \mathcal A^B= \int_{\mathcal S} \mathcal F^B$~we can rewrite Eq. (\ref{eqn:berryhidden}) as
\begin{align}
c_n(T)=c_n(0)\text{e}^{-i\varphi_\gamma^B}.
\label{eqn:berryphase}
\end{align}
The manifestly gauge invariant Berry phase $\varphi_\gamma^B$~can have observable consequences due to interference effects between coherent superpositions that undergo different adiabatic evolutions. The analogue of this phenomenology due to an ordinary electromagnetic vector potential is known as the Aharonov-Bohm effect \cite{AharonovBohm}. The geometrical reason why Berry's connection $\mathcal A^B$~cannot be gauged away all the way along a cyclic adiabatic evolution is the same as why a vector potential cannot be gauged away along a closed path that encloses magnetic flux, namely the notion of holonomy on a curved manifold. We will come back to the concept of holonomy shortly from a more mathematical point of view. For now we only comment that the Berry phase $\varphi_\gamma^B$~is a purely geometrical quantity which only depends on the inner-geometrical relation of the family of states $\lvert n\left(R\right)\rangle$~along the loop $\gamma$~and reflects an abstract notion of curvature in 
Hilbert space which has been defined as Berry's curvature $\mathcal F^B$. 

\subsection{Proof due to Kato}
For a degenerate eigenvalue, Berry's phase is promoted to a unitary matrix acting on the corresponding degenerate eigenspace \cite{WilczekZee1984}. The first proof of the adiabatic theorem of quantum mechanics that overcomes both the limitation to non-degenerate Hamiltonians and the assumption of an explicit phase gauge for the instantaneous eigenstates was reported in the seminal work by Tosio Kato \cite{Kato1950} in 1950. We will review Kato's results briefly for the reader's convenience and use his ideas to illustrate the geometrical origin of the adiabatic phase. The explicit proofs are presented at a very elementary and self contained level in Ref. \onlinecite{Kato1950}. Our notation follows Ref. \onlinecite{AvronQuadrupole1989} which is convenient to relate the physical quantities to elementary concepts of differential geometry.     

Let us assume without loss of generality that the system is at time $t_0=0$~in its instantaneous ground state $\lvert \Psi_0(0)\rangle$~or, more generally, since the ground state might be degenerate, in a state $\lvert \Psi\rangle$~satisfying
\begin{align}
P(0)\lvert \Psi\rangle=\lvert \Psi\rangle,
\label{eqn:projprop}
\end{align}
where $P(t)$~is the projector onto the eigenspace associated with the instantaneous ground state energy $E_0(t)$ which is defined as
\begin{align*}
P(t)=\frac{1}{2\pi i}\oint_c \frac{\text{d}z}{z-\mathcal H(t)},
\end{align*}
where the complex contour $c$~encloses $E_0(t)$~which is again assumed to be separated from the spectrum of excitations by a finite energy gap $\Delta>0$. To understand the adiabatic evolution, we are not interested in the dynamical phase $\phi_D(t)=\int_0^tE_0(\tau)\text{d}\tau$. We thus define a new time evolution operator $\tilde {\mathcal U}(t,0)= \text{e}^{i\phi_D(t)}\mathcal U(t,0)$. Clearly, $\tilde {\mathcal U}$~represents the exact time evolution operator of a system which has the same eigenstates as the original system but has been subjected to a time dependent energy shift that transforms $E_0(t)\rightarrow \tilde E_0(t)=0~\forall_t$. Kato proved the adiabatic theorem in a very constructive way by writing down explicitly the generator $\mathcal A$ of the adiabatic evolution:
\begin{align}
\mathcal A\left(\frac{d}{dt}\right)= -\left[\dot P,P\right].
\label{eqn:Katoconnection}
\end{align}
In the adiabatic limit, $\tilde {\mathcal U}(t,0)P(0)$~was shown \cite{Kato1950} to converge against the adiabatic Kato propagator $\mathcal K$, i.e.,
\begin{align}
\tilde {\mathcal U}(t,0)P(0) \overset{\text{adiabatic limit}}{\longrightarrow} \mathcal K(t,0)=\mathcal T\text{e}^{-\int_0^t\mathcal A\left(\frac{d}{d\tau}\right)\text{d}\tau}.
\label{eqn:Katotheorem}
\end{align}
The adiabatic assumption is now a direct corollary from Eq. (\ref{eqn:Katotheorem}) and can be elegantly expressed as \cite{AvronQuadrupole1989}
\begin{align}
P(t)\mathcal K(t,0)=\mathcal K(t,0) P(0),
\label{eqn:Katoadassum}
\end{align}
implying that a system, which is prepared in an instantaneous ground state at $t_0=0$,~will be propagated to a state in the subspace of instantaneous ground states at $t$~by virtue of Kato's propagator $\mathcal K$. Note that $\mathcal K$~is a completely gauge invariant quantity, i.e., independent of the choice of basis in the possibly degenerate subspace of ground states. The Kato propagator $\mathcal K(T,0)$~associated with a cyclic evolution in parameter space thus yields the Berry phase \cite{Berry} and its non-Abelian generalization \cite{WilczekZee1984}, respectively. We will call this general adiabatic phase the geometric phase (GP) in the following. The GP $\mathcal K_\gamma$~representing the adiabatic evolution along a loop $\gamma$~in parameter space can be expressed in a manifestly gauge invariant way as
\begin{align}
\mathcal K_\gamma = \mathcal T \text{e}^{-\int_\gamma \mathcal A}.
\label{eqn:gp}
\end{align}
Kato's propagator is the solution of an adiabatic analogue of the Schr\"odinger equation, an adiabatic equation of motion that can be written as
\begin{align}
\left(\frac{d}{dt}+\mathcal A\left(\frac{d}{dt}\right)\right)\lvert \Psi(t)\rangle=0,
\label{eqn:adev}
\end{align}
for states satisfying $\quad P(t)\lvert \Psi(t)\rangle=\lvert \Psi(t)\rangle$, i.e., states in the subspace of instantaneous groundstates. Before closing the section, we give a general and at least numerically always viable recipe to calculate the Kato propagator $\mathcal K(t,0)$. We first discretize the time interval $\left[0,t\right]$~into $n$~steps by defining $t_i = i\frac{t}{n}$. The discrete version of Eq. (\ref{eqn:adev}) for the Kato propagator reads (see Eq. (\ref{eqn:Katoconnection}))
\begin{align}
&\mathcal K(t_{i},0)-\mathcal K(t_{i-1},0)=\left\{\left(P(t_{i})-P(t_{i-1})\right)P(t_{i-1})-\right.\nonumber\\
&\left. P(t_i)\left(P(t_{i})-P(t_{i-1})\right)\right\}\mathcal K(t_{i-1},0).
\label{eqn:Katodiscrete}
\end{align}
Using $P(t_{i-1})\mathcal K(t_{i-1},0)=\mathcal K(t_{i-1},0)$~and $P^2=P$, Eq. (\ref{eqn:Katodiscrete}) can be simplified to
\begin{align*}
\mathcal K(t_{i},0)=P(t_i)\mathcal K(t_{i-1},0),
\end{align*}
which is readily solved by $K(t_{i},0)=\prod_{j=0}^iP(t_j)$. Taking the continuum limit yields \cite{Simon1983,WilczekZee1984,AvronQuadrupole1989}
\begin{align}
\mathcal K(t,0)=\lim_{n\to \infty}\prod_{i=0}^nP(t_i),
\label{eqn:Katopropnum}
\end{align}
which is a valuable formula for the practical calculation of the Kato propagator.

\section{Geometric interpretation of adiabatic phases}
\label{sec:adgeo}
We now view the adiabatic time evolution as an abstract notion of parallel transport in Hilbert space and reveal the GP associated with a cyclic evolution as the phenomenon of holonomy due to the presence of curvature in the vector bundle of ground state subspaces over the manifold $\mathcal R$~of control parameters. Interestingly, Kato's approach to the problem provides a gauge invariant, i.e., a global definition of the geometrical entities connection and curvature, whereas standard gauge theories are defined in terms of a complete set of local gauge fields along with their transition functions defined in the overlap of their domains. This difference has an interesting physical ramification: Quantities that are gauge dependent in an ordinary gauge theory like quantum chromodynamics (QCD) are physical observables in the theory of adiabatic time evolution. To name a concrete example, only gauge invariant quantities like 
the trace of the holonomy, also known as the Wilson loop, are observable in QCD whereas the holonomy itself, in other words the GP defined in Eq.(\ref{eqn:gp}), is a physical observable in Kato's theory. This subtle difference has been overlooked in standard literature on this subject \cite{ZeeNuclearBerry1988,BohmBerryBook} which we interpreted as an incentive to clarify this point below in greater detail.

\subsection{Adiabatic time evolution and parallel transport}
To get accustomed to parallel transport, we first explain the general concept with the help of a very elementary example, namely a smooth piece of two dimensional surface embedded in $\mathbb R^3$. If the surface is flat, there is a trivial notion of parallel transport of tangent vectors, namely shifting the same vector in the embedding space from one point to another. However, on a curved surface, this program is ill-defined, since a tangent vector at one point might be the normal vector at another point of the surface. Put shortly, a tangent vector can only be transported as parallel as the curvature of the surface admits. On a curved surface, parallel transport along a curve is thus defined as a vanishing in-plane component of the directional derivative, i.e., a vanishing covariant derivative of a vector field along a curve. The normal component of the directional derivative reflects the rotation of the entire tangent plane in the embedding 
space and is not an inner-geometric quantity of the surface as a two dimensional manifold.\\

The analogue of the curved surface in the context of adiabatic time evolution is the manifold of control parameters $\mathcal R$, parameterizing for example external magnetic and electric fields. The analogue of the tangent plane at each point of the surface is the subspace of degenerate ground states of the Hamiltonian $\mathcal H(R)$~at each point $R$~in parameter space. An adiabatic time dependence of $\mathcal H$~amounts to traversing a curve $t\mapsto R(t)$~in $\mathcal R$~at adiabatically slow velocity. A cyclic evolution is uniquely associated with a loop $\gamma$~in $\mathcal R$. We will now explicitly show that the adiabatic equation of motion (\ref{eqn:adev}) defines a notion of parallel transport in the fiber bundle of ground state subspaces over $\mathcal R$~in a completely analogous way as the ordinary covariant derivative $\nabla$~on a smooth surface defines parallel transport in the tangent bundle of the smooth surface. We first note that $\frac{d}{dt}=\dot R^\mu \partial_\mu$~is referring 
to a particular direction $\dot R^\mu$~in parameter space, which depends on the choice of the adiabatic time dependence of $\mathcal H$. We can get rid of this dependence by rephrasing Eq. (\ref{eqn:adev}) as
\begin{align}
D \lvert \Psi\rangle=\left(d+\mathcal A\right) \lvert \Psi\rangle=0,
\label{eqn:covd}
\end{align}
where the adiabatic derivative $D=d+\mathcal A$~has been defined, $\mathcal A = -\left[(dP),P\right]$~and here as in the following $P\lvert \Psi\rangle=\lvert \Psi\rangle$. The $R$-dependence has been dropped for notational convenience. The adiabatic derivative $D$~takes a tangent vector, e.g., $\frac{d}{dt}$, as an argument to boil down to the directional adiabatic derivative $\frac{d}{dt}+\mathcal A\left(\frac{d}{dt}\right)$~appearing in Eq. (\ref{eqn:adev}). For the following analysis the identities $P^2=P$~and $P\lvert \Psi\rangle=\lvert \Psi\rangle$~are of key importance. It is now elementary algebra to show
\begin{align}
P(dP)P=0.
\label{eqn:katoid}
\end{align}
Eq. (\ref{eqn:katoid}) has a simple analogue in elementary geometry: Consider the family of unit vectors $\left\{n(t)\right\}_t$~where $t$~parameterizes a curve on a smooth surface. Then, since $1= \langle n\vert n\rangle$, we get $0= \frac{d}{dt}\langle n\vert n\rangle= 2 \langle n\vert\dot n\rangle$, i.e., the change of a unit vector is perpendicular to the unit vector itself.
Using Eq. (\ref{eqn:katoid}), we immediately derive $P\mathcal A \lvert \Psi \rangle=0$~and with that
\begin{align}
D \lvert \Psi\rangle=0 \Leftrightarrow Pd \lvert \Psi\rangle=0.
\label{eqn:geoint}
\end{align}
This makes the analogy of our adiabatic derivative $D$~to the ordinary notion of parallel transport manifest: $\lvert \Psi\rangle$~is parallel-transported if the in-plane component of its derivative vanishes.   

\subsubsection*{Curvature and holonomy}

Let us again start with a very simple example of a curved manifold, a two dimensional sphere $S^2$, which has constant Gaussian curvature. Parallel-transporting a tangent vector around a geodesic triangle, say the boundary of an octant of the sphere gives a defect angle which is proportional to the area of the triangle or, more precisely, the integral of the Gaussian curvature over the enclosed area. This defect angle is called the holonomy of the traversed closed path. This elementary example suggests that the presence of curvature is in some sense probed by the concept of holonomy. This intuition is absolutely right. As a matter of fact, the generalized curvature at a given point $x$~of the base manifold of a fiber bundle is defined as the holonomy associated with an infinitesimal loop at $x$. More concretely, the curvature $\Omega$~is usually defined as $\Omega_{\mu\nu}=\left[\nabla_\mu,\nabla_\nu\right]$~which represents an infinitesimal parallel transport around a parallelogram in the $\mu\nu$-plane.\\

In total analogy, we define
\begin{align}
\mathcal F_{\mu\nu}\lvert \Psi\rangle= \left[D_\mu,D_\nu\right]\lvert\Psi\rangle= P\left[P_\mu,P_\nu\right]P\lvert \Psi\rangle,
\label{eqn:curvconstruct}
\end{align}
with the shorthand notation $P_\mu=\partial_\mu P$. Restricting the domain of $\mathcal F$~to states which are in the projection $P$, we can rewrite Eq. (\ref{eqn:curvconstruct}) as the operator identity
\begin{align}
\mathcal F = \mathcal F_{\mu\nu}dR^\mu\wedge dR^\nu=P\left[(dP)\wedge(dP)\right]P.
\label{eqn:Katocurv}
\end{align}

In the general case of a non-Abelian adiabatic connection, i.e., if the dimension of $P$~is larger than $1$, we cannot simply use Stokes theorem to reduce the evaluation of Eq. (\ref{eqn:gp}) to a surface integral of $\mathcal F$~over the surface bounded by $\gamma$, as has been done in the case of the Abelian Berry curvature in Eq. (\ref{eqn:StokesBerry}). However, the global one to one correspondence between curvature and holonomy still exists and is the subject of the Ambrose-Singer theorem \cite{Nakahara}.

\subsubsection*{Relation between Kato's and Berry's language}

In order to make contact to the more standard language of gauge theory, we will now express Kato's manifestly gauge invariant formulation \cite{Kato1950} in local coordinates thereby recovering Berry's connection $\mathcal A^B$~\cite{Berry,Simon1983} and its non-Abelian generalization \cite{WilczekZee1984}, respectively. For this purpose, let us fix a concrete basis $\left\{\lvert \alpha(R)\rangle\right\}_\alpha$, $R\in O\subset \mathcal R$~in an open subset $O$~of the parameter manifold. We assume the loop $\gamma$~to lie inside of $O$.
Otherwise we would have to switch the gauge while traversing the loop. We will drop the $R$-dependence of $\lvert \alpha\rangle$~right away for notational convenience. The projector $P$~can then be represented as $P=\sum_\alpha \lvert \alpha\rangle\langle \alpha\rvert$. Let us start the cyclic evolution without loss of generality with $\lvert \Psi(0)\rangle=\lvert \alpha(0)\rangle$. From Eq. (\ref{eqn:Katoadassum}) we know that the solution $\lvert \Psi(t)\rangle=\mathcal K(t,0)\lvert\alpha(0)\rangle$~of Eq. (\ref{eqn:adev}) satisfies $P(t)\lvert \Psi(t)\rangle=\lvert \Psi(t)\rangle$~at every point in time during the cyclic evolution. Hence, we can represent $\lvert \Psi(t)\rangle$~in our gauge as
\begin{align}
\lvert \Psi\rangle= \sum_\beta \langle \beta\rvert \Psi\rangle\lvert \beta\rangle= U^B_{\beta\alpha}\lvert \beta\rangle,
\label{eqn:berrygauge}
\end{align}
where the $t$-dependence has been dropped for brevity. From Eq. (\ref{eqn:geoint}), we know that $P\frac{d}{dt}\lvert\Psi\rangle=0$~which implies $\langle \gamma\rvert\frac{d}{dt}\lvert\Psi\rangle=0$. Plugging this into Eq. (\ref{eqn:berrygauge}) yields
\begin{align}
\frac{d}{dt}U^B_{\gamma \alpha}=-\sum_\beta \langle \gamma\rvert \frac{d}{dt}\lvert \beta\rangle U^B_{\beta \alpha}.
\label{eqn:berryu}
\end{align}
Redefining $\mathcal A^B$~for the non-Abelian case as a matrix valued gauge field through $\mathcal A^B_{\alpha\beta}= -i\langle \alpha\rvert\partial_\mu\lvert \beta\rangle dR^\mu$, Eq. (\ref{eqn:berryu}) is readily solved as
\begin{align*}
U^B(t)=\mathcal T \text{e}^{-i\int_0^t \mathcal A^B(\frac{d}{d\tau})d\tau}.
\end{align*}
The representation matrix of the GP associated with the loop $\gamma$~then reads
\begin{align}
U^B_\gamma =\mathcal T \text{e}^{-i\int_\gamma \mathcal A^B}.
\label{eqn:berryholonomy}
\end{align}
By construction, $U^B_\gamma$~is the representation matrix of the GP $\mathcal K_\gamma$, i.e.,
\begin{align*}
\left(U^B_\gamma\right)_{\alpha,\beta}=\langle \alpha(0)\rvert \mathcal K_\gamma\lvert \beta(0)\rangle,
\end{align*}
or, more general, for any point in time along the path
\begin{align}
\left(U^B(t)\right)_{\alpha,\beta}=\langle \alpha(t)\rvert \mathcal K(t,0)\lvert \beta(0)\rangle.
\label{eqn:relationKatoBerry}
\end{align}
Eq. (\ref{eqn:relationKatoBerry}) makes the relation between Kato's formulation of adiabatic time evolution and the non-Abelian Berry phase manifest. In contrast to the gauge independence of Kato's global connection $\mathcal A$, $\mathcal A^B$~behaves like a local connection (see Ref. \onlinecite{ChoquetBruhat} for rigorous mathematical definitions) and depends on the gauge, i.e., on our choice of the family $\left\{\lvert\alpha(R)\rangle\right\}_\alpha$~of basis states. Under a smooth family of basis transformations $\left\{U(R)\right\}_R$~acting on the local coordinates $\mathcal A^B$~transforms like \cite{ChoquetBruhat,Nakahara}  
\begin{align}
\mathcal A^B\rightarrow \tilde{\mathcal A}^B= U^{-1}\mathcal A^BU+U^{-1}dU
\label{eqn:aberrygaugetrans}
\end{align}
resulting in the following gauge dependence of Eq. (\ref{eqn:berryholonomy}),
\begin{align}
U^B_\gamma\rightarrow \tilde U^B_\gamma= U^{-1}U^B_\gamma U,
\label{eqn:gaugedephol}
\end{align}
which only depends on the basis choice $U= U\left(R(0)\right)$~at the starting point of the loop $\gamma$.\\

Inserting our representation $P=\sum_\alpha \lvert\alpha\rangle\langle \alpha\rvert$~into the gauge independent form of the curvature, Eq.(\ref{eqn:Katocurv}), we readily derive
\begin{align*}
\mathcal F^{B,\alpha\beta}_{\mu\nu}= \langle \alpha \rvert \left[P_\mu,P_\nu\right]\lvert \beta\rangle=(d\mathcal A^{B})_{\mu\nu}^{\alpha\beta}+(\mathcal A^B\wedge A^B)_{\mu\nu}^{\alpha\beta},
\end{align*}
which defines $\mathcal F^B$~as the usual curvature of a non-Abelian gauge field \cite{Nakahara}, i.e.,
\begin{align}
\mathcal F^B=d\mathcal A^B+\mathcal A^B\wedge \mathcal A^B,
\label{eqn:nonabBerrycurv}
\end{align}
which transforms under a local gauge transformation $U$~like
\begin{align*}
\mathcal F^B\rightarrow U^{-1}\mathcal F^B U. 
\end{align*}

\subsection{Gauge dependence and physical observability}
\label{sec:observability}

The gauge dependence of the non-Abelian Berry phase $U^B_\gamma$~(see Eq. (\ref{eqn:gaugedephol})) has led several authors \cite{ZeeNuclearBerry1988,BohmBerryBook} to the conclusion that only gauge independent features like the trace and the determinant of $U^B_\gamma$~can have physical meaning. However, working with Kato's manifestly gauge invariant formulation, it is understood that the entire GP $\mathcal K_\gamma$~is experimentally observable. In the remainder of this section we will try to shed some light on this ostensible controversy.\\

In gauge theory, it goes without saying that explicitly gauge dependent phenomena are not immediately physically observable and that only the gauge invariant information resulting from a calculation performed in a special gauge can be of physical significance. At a formal level this is a direct consequence of the fact that the Lagrangian of a gauge theory is constructed in a manifestly gauge invariant way by tracing over the gauge space indices. The physical reason for this is quite simple: A concrete gauge amounts to a local choice of the coordinate system in the gauge space. Under a local change of basis, a non-abelian gauge field $A$~transforms like (see also Eq. (\ref{eqn:aberrygaugetrans}))  
\begin{align*}
A\rightarrow \tilde A= U^{-1}AU+U^{-1}dU, 
\end{align*}
where $U(x)$~is a smooth family of basis transformations, with $x$~labeling points in the base space of the theory, e.g., in Minkowski space. Now, since the gauge space is an internal degree of freedom, the basis vectors in this space are not associated with physical observables. This situation is fundamentally changed in Kato's adiabatic analogue of a gauge theory. Here, the non-Abelian structure is associated with a degeneracy of the Hamiltonian, e.g., Kramers degeneracy in the presence of time reversal symmetry (TRS). For a system in which spin is a good quantum number, Kramers degeneracy is just spin degeneracy, which makes the spin the analogue of the gauge degree of freedom in an ordinary gauge theory. However, the magnetic moment associated with a spin is a physical observable which can be measured. The basis vectors, e.g.,  $\lvert \uparrow \rangle,\lvert \downarrow\rangle$~have an objective meaning for the experimentalist (a magnetic moment that points from the lab-floor to the sky which we call $z$-
direction). For concreteness, let us assume that we have calculated a GP $\mathcal K_\gamma=\lvert \uparrow \rangle\langle \downarrow\rvert+\lvert\downarrow \rangle\langle \uparrow\rvert$. The representation matrix of $\mathcal K_\gamma$~in this basis of $S_z$~eigenstates is clearly the Pauli matrix $\sigma_x$. Choosing a different gauge, i.e., a different basis for the gauge degree of freedom at the starting point of the cyclic adiabatic evolution, we of course would have obtained a different representation matrix $U^B_\gamma$ for $\mathcal K_\gamma$, e.g., $\sigma_z$, had we chosen the basis as eigenstates of $S_x$ (see Eq. (\ref{eqn:gaugedephol})). However, the fact that $\mathcal K_\gamma$~rotates a spin which is initially pointing to the lab-ceiling upside down is gauge independent physical reality.

\section{From geometry to topology}
\label{sec:tsmintro}
In Section \ref{sec:adgeo}, we worked out the relation between the GP and the notion of curvature as a local geometric quantity. The topological invariants characterizing TSM are in some sense global GPs. They measure global properties which cannot be altered by virtue of local continuous changes of the physical system. Continuous is at this stage of the analysis synonymous with adiabatic, i.e., happening at energies below the bulk gap. Later on, we will additionally require local continuous changes to respect the fundamental symmetries of the physical system, e.g., particle hole symmetry (PHS) or time reversal symmetry (TRS).\\

Let us illustrate the correspondence between local curvature and global topology of a manifold with the help of the simplest possible example. We consider a two dimensional sphere $S^2$~with radius $r$. This manifold has a constant Gaussian curvature of $\kappa= \frac{1}{r^2}$. The integral of $\kappa$~over the entire sphere obviously gives $4\pi$, independent of $r$. The Gauss-Bonnet theorem in its classical form (see, e.g., Ref. \onlinecite{Kuehnel2005}) relates precisely this integral of the Gaussian curvature of a closed smooth two dimensional manifold $\mathcal M$~to its Euler characteristic $\chi$~in the following way:
\begin{align}
\frac{1}{2\pi}\int_{\mathcal M}\kappa = \chi(\mathcal M)
\label{eqn:GaussBonnetClass}
\end{align}
Note that $\chi$~is a purely algebraic quantity which is defined as the number of vertices minus the number of edges plus the number of faces of a triangulation \cite{NashSen} of $\mathcal M$. $\chi$~is by construction of simplicial homology \cite{NashSen} a topological invariant which can only be changed by poking holes into $\mathcal M$~and gluing the resulting boundaries together so as to create closed manifolds with different genus. Hence, Eq. (\ref{eqn:GaussBonnetClass}) nicely demonstrates how the integral of the local inner-geometric quantity $\kappa$~over the entire manifold yields a topological invariant. Concretely, for our example $S^2$, a triangulation is provided by continuously deforming the sphere into a tetrahedron. Simple counting of vertices, edges, and faces yields $\chi(S^2)=4-6+4=2$, in agreement with Eq. (\ref{eqn:GaussBonnetClass}). More generally speaking, $e=\frac{\kappa}{2\pi}$~is our first encounter with a characteristic class \cite{MilnorStasheff}, the so called Euler class of $\mathcal M$, which upon integration over $\mathcal M$~yields the topological invariant $\chi$. Similar mathematical structures will be ubiquitous when it comes to the classification of TSM. The simplest example in this context is the quantum anomalous Hall (QAH) state \cite{QAH,QAHPRB,QAHShoucheng}, a 2D insulating state in symmetry class A (see Section \ref{sec:caz}) which is characterized by its first Chern number (see Section \ref{sec:complexclasses} for a detailed discussion)
\begin{align}
\mathcal C_1=\int_{\text{BZ}} \frac{i\text{Tr}\left[\mathcal F\right]}{2\pi}.
\label{eqn:firstchern}
\end{align}
The formal analogy between Eq. (\ref{eqn:GaussBonnetClass}) and Eq. (\ref{eqn:firstchern}) is striking. The parameter space over which the adiabatic curvature $\mathcal F$~is integrated in Eq. (\ref{eqn:firstchern}) is the $k$-space of the physical system, i.e., the Brillouin zone (BZ) for a periodic system, which has the topology of a torus. In Section \ref{sec:adtheo}, we showed that the GP can be viewed as the flux through the surface in parameter space which is bounded by the corresponding cyclic adiabatic evolution. Along similar lines, the first Chern number is analogous to a monopole charge enclosed by the entire BZ of the QAH insulator.\\

The close correspondence between adiabatic evolution and the first Chern number has also been discussed in Refs. {\cite{Laughlin1981,ZakPol,FuPump}: For a 2D system in cylinder geometry, a circumferential electric field can be modeled by a time dependent magnetic flux threading the cylinder in axial direction. In this scenario, the first Chern number can be viewed as the shift of the charge polarization in axial direction associated with the adiabatic threading of one flux quantum. Along these lines, Laughlin \cite{Laughlin1981} had interpreted the quantum Hall effect as an adiabatic charge pumping process, even before the formal relation between the Hall conductivity and the first Chern number was established in Refs. \onlinecite{TKNN1982,Avron1983,Kohmoto1985}.

\section{Bulk classification of all non-interacting TSM}

\label{sec:symmclass}
Very generally speaking, the understanding of TSM can be divided into two subproblems. First, finding the group that represents the topological invariant for a class of systems characterized by their fundamental symmetries and spatial dimension. Second, assigning the value of the topological invariant to a representative of such a symmetry class, i.e., measuring to which topological equivalence class a given system belongs. We address the first problem in this section and the second problem 
in \ref{sec:topcalc}.\\

The general idea that yields the entire table of TSM is quite simple: In addition to requiring a bulk insulating gap, physical systems of a given spatial dimension are divided into 10 symmetry classes distinguished by their fundamental symmetries, i.e., TRS, PHS, and chiral symmetry (CS) \cite{AltlandZirnbauer}. The topological properties of the corresponding Cartan symmetric spaces of quadratic candidate Hamiltonians determine the group of possible topologically inequivalent systems. We outline the mathematical structure behind this general classification scheme in some detail. First, we briefly review the construction of the ten universality classes \cite{AltlandZirnbauer}. Then, we present the associated topological invariants for non-interacting systems of arbitrary spatial dimension giving a complete list of all TSM \cite{RyuLudwig} that can be distinguished by virtue of this framework. Finally, we discuss in some detail the origin of characteristic patterns appearing in this table using the 
framework of K-Theory along the lines of the pioneering work by Kitaev \cite{KitaevPeriodic}.  

\subsection{Cartan-Altland-Zirnbauer symmetry classes}
\label{sec:caz}
A physical system can have different types of symmetries. An ordinary symmetry \cite{RyuLudwig} is characterized by a set of unitary operators representing the symmetry operations that commute with the Hamiltonian. The influence of such a symmetry on the topological classification can be eliminated by transforming the Hamiltonian into a block-diagonal form with symmetry-less blocks. The total system then consists of several uncoupled copies of symmetry-irreducible subsystems which can be classified individually. In contrast, the ``extremely generic symmetries'' \cite{RyuLudwig} follow from the anti-unitary operations of TRS and PHS. Involving complex conjugation according to Wigner, they impose certain reality conditions on the system Hamiltonian. In total, the behavior of the system under these operations, and their combination, the CS operation, defines ten universality classes which we call the Cartan-Altland-Zirnbauer (CAZ) classes. For disordered systems, these classes correspond to ten distinct 
renormalization group (RG) low energy fixed points in random matrix theory \cite{AltlandZirnbauer}. The spaces of candidate Hamiltonians within these symmetry classes correspond to the ten symmetric spaces introduced by Cartan in 1926 \cite{Cartan} defined in terms of quotients of Lie groups represented in the Hilbert space of the system. For translation-invariant systems, the imposed reality conditions are inherited by the Bloch Hamiltonian $h(k)$~(see Eq. (\ref{eqn:symBloch}) below).\\

In the following, the anti-unitary TRS operation will be denoted by $\mathcal T$~and the anti-unitary PHS will be denoted by $\mathcal C$. The Hamiltonian $\mathcal H$~of a physical system satisfies these symmetries if
\begin{align}
\mathcal T \mathcal H\mathcal T^{-1}=\mathcal H,
\label{eqn:trsop}
\end{align}
and
\begin{align}
\mathcal C \mathcal H\mathcal C^{-1}=-\mathcal H,
\label{eqn:phsop}
\end{align}
respectively. According to Wigner's theorem, these anti-unitary symmetries can be represented as a unitary operation times the complex conjugation $K$. We define $\mathcal T = TK,~\mathcal C=CK$. Using the unitarity of $T,C$~along with $\mathcal H = \mathcal H^\dag$~we can rephrase Eqs. (\ref{eqn:trsop}-\ref{eqn:phsop}) as
\begin{align}
& T\mathcal H^TT^\dag=\mathcal H,\nonumber\\
& C\mathcal H^TC^\dag=-\mathcal H.
\label{eqn:symmat}
\end{align}
There are two inequivalent realizations of these anti-unitary operations distinguished by their square which can be plus identity or minus identity. For example, $\mathcal T^2=\pm 1$~for the unfolding of a particle with integer/half-integer spin, respectively. Clearly, $\mathcal T^2=\pm 1\Leftrightarrow TT^*=\pm 1$~and $\mathcal C^2=\pm 1\Leftrightarrow CC^*=\pm 1$. In total, there are thus nine possible ways for a system to behave under the two anti-unitary symmetries: each symmetry can be absent, or present with square plus or minus identity. For eight of these nine combinations, the behavior under the combination $\mathcal T\mathcal C$~is fixed. The only exception is the so called unitary class which breaks both PHS and TRS and can either obey or break their combination, the CS. This class hence splits into two universality classes which add up to a grand total of ten classes shown in Tab. \ref{tab:CAZ}.
\begin{table}[h]
\centering
\begin{tabular}{|l|ccc|}
 \hline
 Class &    TRS    & PHS&CS  \\ \hline
 A~(Unitary)& 0&0&0\\
 AI~(Orthogonal)&+1&0&0\\
 AII~(Symplectic)&-1&0&0\\ \hline
 AIII~(Chiral Unitary)&0&0&1\\
 BDI~(Chiral Orthogonal)&+1&+1&1\\
 CII~(Chiral Symplectic)&-1&-1&1\\ \hline
 D&0&+1&0\\
 C&0&-1&0\\
 DIII&-1&+1&1\\
 CI&+1&-1&1\\ \hline
\end{tabular}
\caption{\label{tab:CAZ}
Table of the CAZ universality classes. 0 denotes the absence of a symmetry. For PHS and TRS, $\pm$1 denotes the square of a present symmetry, the presence of CS is denoted by 1. The last four classes are Bogoliubov de$\,$Gennes classes of mean field superconductors where the superconducting gap plays the role of the insulating gap.
}
\end{table}
For a periodic system, symmetry constraints similar to Eq. (\ref{eqn:symmat}) hold for the Bloch Hamiltonian $h(k)$, namely
\begin{align}
& T h^T(-k)T^\dag=h(k),\nonumber\\
& C h^T(-k)C^\dag=-h(k),
\label{eqn:symBloch}
\end{align}
where $T,C$~now denote the representation of the unitary part of the anti-unitary operations in band space.\\

For a continuum model, the real space Hamiltonian $H(x)$~is defined through
\begin{align*}
\mathcal H = \int \text{d}^dx~\Psi^\dag(x)H(x)\Psi(x),
\end{align*}
where $\Psi$~is a vector/spinor comprising all internal degrees like spin, particle species, etc.. The $k$-space on which the Fourier transform $\tilde H(k)$~of $H(x)$~is defined does not have the topology of a torus like the BZ of a periodic system. However, the continuum models one is concerned with in condensed matter physics are effective low energy/large distance theories.
For large $k$, $\tilde H(k)$~will thus generically have a trivial structure so that the $k$-space can be endowed with the topology of the sphere $S^d$~by a one point compactification which maps $k\rightarrow\infty$~to a single point. The symmetry constraints on $\tilde H(k)$~have the same form as those on the Bloch Hamiltonian $h(k)$~shown in Eq. (\ref{eqn:symBloch}). By abuse of notation, we will denote both $\tilde H(k)$~and $h(k)$~by $h(k)$. Nevertheless, we will point out several differences between periodic systems and continuum 
models along the way.  

\subsection{Definition of the classification problem for continuum models and periodic systems}
For translation-invariant insulating systems with $n$~occupied and $m$~empty bands and continuum models with $n$~occupied and $m$~empty fermion species, respectively, the projection $P(k)= \sum_{\alpha=1}^n\lvert u_\alpha(k)\rangle\langle u_\alpha(k)\rvert$~onto the occupied states is the relevant quantity for the topological classification. The spectrum of the system is not of interest for adiabatic quantities as long as a bulk gap between the empty and the occupied states is maintained. We thus deform the system adiabatically into a flat band insulator, i.e., a system with eigenenergy $\epsilon_-=-1$~for all occupied states and eigenenergy  $\epsilon_+=+1$~for all empty states. The eigenstates are not changed during this deformation. The Hamiltonian of this flat band system then reads \cite{QiTFT,Schnyder2008}
\begin{align*}
Q(k)= (+1)\left(1-P(k)\right)+(-1)P(k)=1-2P(k) 
\end{align*}
Obviously, $Q^2=1, \text{Tr}\left[Q\right]=m-n$. Without further symmetry constraints, $Q$~is an arbitrary $U(n+m)$~matrix which is defined up to a $U(n)\times U(m)$~gauge degree of freedom corresponding to basis transformations within the subspaces of empty and occupied states, respectively. Thus, $Q$~is in the symmetric space
\begin{align*}
G_{n+m,m}(\mathbb C)=G_{n+m,n}(\mathbb C)=U(n+m)/(U(n)\times U(m)).
\end{align*}
Geometrically, the complex Grassmannian $G_{k,l}(\mathbb C)$~is a generalization of the complex projective plane and is defined as the set of $l$-dimensional planes through the origin of $\mathbb C^k$. The set of topologically different translation-invariant insulators is then given by the group $g$~of homotopically inequivalent maps $k\mapsto Q(k)$~from the BZ $T^d$~of a system of spatial dimension $d$~to the space $G_{n+m,m}(\mathbb C)$ of possible Bloch Hamiltonians. For continuum models $T^d$~is replaced by $S^d$ and $g$~is by definition given by
\begin{align}
g = \pi_d\left(G_{n+m,m}(\mathbb C)\right),
\label{eqn:purehomotopy}
\end{align}
where the $n$-th homotopy group $\pi_n$~of a space is by definition the group of homotopically inequivalent maps from $S^d$~to this space. For translation-invariant systems defined on a BZ, the classification can be more complicated than Eq. (\ref{eqn:purehomotopy}) if the lower homotopy groups $\pi_s,~s=1,\ldots, d-1$~are nontrivial. For the quantum anomalous Hall (QAH) insulator \cite{QAH}, a 2D translation-invariant state which does not obey any fundamental symmetries, we can infer from $\pi_2\left(G_{n+m,m}(\mathbb C)\right)=\mathbb Z,~\pi_1\left(G_{n+m,m}(\mathbb C)\right)={0}$~that an integer topological invariant must distinguish possible states of matter in this symmetry class, i.e., possible maps $T^2\rightarrow G_{n+m,m}(\mathbb C)$.
The condition  $\pi_1\left(G_{n+m,m}(\mathbb C)\right)={0}$~is necessary because the $\pi_2$~classifies maps from $S^2$, which is only equivalent to the classification of physical maps from $T^2$~if the fundamental group $\pi_1$~of the target space is trivial. The 
difference between the base space of a periodic systems which is a torus and of continuum models which has the topology of a sphere has interesting physical ramifications: The so called weak topological insulators are only topologically distinct over a torus but not over a sphere. Physically, this is visible in the lacking robustness of these TSM which break down with the breaking of translation symmetry.\\

Requiring further symmetries as appropriate for the other nine CAZ universality classes is tantamount to imposing symmetry constraints on the allowed maps $T^d\rightarrow G_{n+m,m}(\mathbb C),~k\mapsto Q(k)$~for translation-invariant systems and $S^d\rightarrow G_{n+m,m}(\mathbb C),~k\mapsto Q(k)$~for continuum models, respectively. The set of topologically distinct physical systems is then still given by the set of homotopically inequivalent maps within this restricted space, i.e., the space of maps which cannot be continuously deformed into each other without breaking a symmetry constraint. For example, for the chiral classes characterized by $\text{CS}=1$, $Q$~can be brought into the off diagonal form \cite{Schnyder2008}
\begin{align*}
Q=\begin{pmatrix}0&q\\q^\dag & 0\end{pmatrix}
\end{align*}
with $q q^\dag = 1$, which reduces the corresponding target space to $U(n)$. For the chiral unitary class AIII without further symmetry constraints, the calculation of $g$~amounts to calculating $g= \pi_d\left(U(n)\right)=\mathbb Z$~for odd $d$~and $g=\pi_d\left(U(n)\right)={0}$~for even $d$, respectively, provided $n \ge (d+1)/2$. Additional symmetries will again impose additional constraints on the map $T^d\rightarrow U(n),~k\mapsto q(k)$.\\

This procedure rigorously defines the group $g$~of topological equivalence classes for non-interacting translation-invariant insulators in arbitrary spatial dimension and CAZ universality class. However, the practical calculation of $g$~can be highly non-trivial and has been achieved for continuum models via various subtle detours, for example the investigation of surface nonlinear $\sigma$-models, in Refs. \onlinecite{SchnyderRyu,RyuLudwig}. In the following, we outline a mathematical brute force solution to the classification problem in terms of K-Theory which has been originally introduced in the seminal work by Kitaev in 2009 \cite{KitaevPeriodic}. This method will naturally explain the emergence of weak topological insulators. We will not assume any prior knowledge on K-Theory. In Tab. \ref{tab:classclean}, we summarize the resulting groups of topological sectors $g$~for all possible systems \cite{SchnyderRyu}. We notice an interesting diagonal pattern relating subsequent symmetry classes to neighboring 
spatial dimensions. Furthermore, the pattern of the two unitary classes shows a periodicity of two in the spatial dimension. If we had shown the classification for higher spatial dimensions, we would have observed a periodicity of eight in the spatial dimension for the eight real classes. The following discussion is dedicated to provide a deeper understanding of these fundamental patterns as pioneered in Ref. \onlinecite{KitaevPeriodic}. The mentioned periodicities have first been pointed out in Refs. \onlinecite{QiTFT,SchnyderRyu}.   
\begin{table*}[ht]
\centering
\begin{tabular}{|l|l|cccc|}\hline
 Class &    constraint    & $d=1$&$d=2$&$d=3$&$d=4$\\ \hline
 A & none&0&$\mathbb Z$&0&$\mathbb Z$\\
 AIII &none on $q$&$\mathbb Z$&0&$\mathbb Z$&0\\ \hline
 AI &$Q^T(k)=Q(-k)$&0&0&0&$\mathbb Z$\\
 BDI&$q^*(k)=q(-k)$&$\mathbb Z$&0&0&0\\
 D&$\tau_x Q^T(k)\tau_x=-Q(-k),~m=n$&$\mathbb Z_2$&$\mathbb Z$&0&0\\
 DIII&$q(k)^T=-q(-k), m=n~\text{even}$&$\mathbb Z_2$&$\mathbb Z_2$&$\mathbb Z$&0\\
 AII &$i\sigma_y Q^T(k)(-i\sigma_y)=Q(-k),~m,n~\text{even}$&0&$\mathbb Z_2$&$\mathbb Z_2$&$\mathbb Z$\\ 
 CII&$i\sigma_y q^*(k)(-i\sigma_y)=q(-k),~m=n~\text{even}$&$\mathbb Z$&0&$\mathbb Z_2$&$\mathbb Z_2$\\ 
 C&$\tau_y Q^T(k)\tau_y=-Q(-k),~m=n$&0&$\mathbb Z$&0&$\mathbb Z_2$\\
 CI&$q(k)^T=q(-k), m=n$&0&0&$\mathbb Z$&$0$\\ \hline
\end{tabular}
\caption{\label{tab:classclean} Table of all groups $g$ of topological equivalence classes. The first column denotes the CAZ symmetry class, divided into two unitary classes without anti-unitary symmetry (top) and eight ``real'' classes with at least one anti-unitary symmetry (bottom). The second column shows the symmetry constraints on the flat band maps, where we have chosen the representation $\mathcal T=K$~for $\mathcal T^2=1$, $\mathcal T=i\sigma_y K$~for $\mathcal T^2=-1$,~as well as $\mathcal C=\tau_x K$~for $\mathcal C^2=1$, $\mathcal C=\tau_y K$~for $\mathcal C^2=-1$. Here, $\sigma_y$~denotes the Pauli matrix in spin space, $\tau_x,\tau_y$~denote Pauli matrices in the particle hole pseudo spin space of Bogoliubov de$\,$Gennes Hilbert spaces. In the last four columns, $g$~is listed for $d=1,\ldots,4$.
}
\end{table*}

\subsection{Topological classification of unitary vector bundles}
In order to prepare the reader for the application of K-Theory and to motivate its usefulness, we first formulate the classification problem in the language of fiber bundles.
The mathematical structure of a non-interacting insulator of spatial dimension $d$~is that of a vector bundle $E\overset{\pi}{\rightarrow} \mathcal M$~(see Ref. \onlinecite{EguchiReview} for a pedagogical review on vector bundles in physics). Roughly speaking, a vector bundle consists of a copy of a vector space $V$~which is called the typical fiber over each point of its so called base manifold $\mathcal M$. The projection $\pi$~of the bundle projects the entire fiber $V_p\simeq V$~over each point $p\in \mathcal M$~to $p$. Conversely, $V_p$~can be viewed as the inverse image of $p$~under $\pi$, i.e. , $V_p=\pi^{-1}\left[p\right]$. The base manifold $\mathcal M$~of $E$ is the $d$~dimensional $k$-space of the system, and the fiber over a point $k$~given by the (projective) space of occupied states $P(k)$. The gauge group of the bundle is $U(n)$, where $n$~is the dimension of $P$. Locally, say in a neighborhood $\mathcal U\subset \mathcal M$~of a given point $p\in \mathcal M$, any vector bundle looks like the trivial product $\mathcal U\times V$. Globally however, this product can be twisted which gives rise to topologically distinct bundles.
If no further symmetry conditions are imposed (see class A in Tab.~\ref{tab:CAZ}), the question of how many topologically distinct insulators in a given dimension exist is tantamount to asking how many homotopically different $U(n)$~vector bundles can be constructed over $\mathcal M$. This question can be formally answered for arbitrary smooth manifolds $\mathcal M$~as we will outline now. The general idea is the following. There is a universal bundle $\xi\overset{\Pi}{\rightarrow}\mathcal X$~into which every bundle $E$~can be embedded through a bundle map \cite{Nakahara} $\hat f: E\rightarrow \xi$~such that
\begin{align}
f^*\xi=E,
\label{eqn:unipullback}
\end{align}
where $f:\mathcal M\rightarrow \mathcal X$~is the map between the base manifolds associated with the bundle map $\hat f$. 
That is to say every bundle can be represented as a pullback bundle $f^*\xi$~\cite{Nakahara} of the universal bundle $\xi$~by virtue of a suitable bundle map $\hat f$. The key point is now that homotopically different bundles $E$~are distinguished by homotopically distinct maps $f$. Thus, the set of different TSM is the set $\pi\left[\mathcal M, \mathcal X\right]$~of homotopy classes of maps from the $k$-space $\mathcal M$~to the base manifold $\mathcal X$~of the universal bundle. $\mathcal X$~is also called the classifying space of $U(n)$~and is given  by the Grassmanian $G_{N,n}(\mathbb C)=U(N)/(U(n)\times U(N-n))$~for sufficiently large $N$, i.e., $N> \lceil\frac{d}{2}+n\rceil$. To be generic in the dimension of the system $d$, we take the inductive limit $\mathcal X=G_{n}(\mathcal C^\infty) = \lim_{N\to \infty} G_{N,n}(\mathbb C)=\lim_{N\to \infty} U(N)/(U(n)\times U(N-n))$. We thus found for the set $\text{Vect}_n(\mathcal M, \mathbb C)$~of inequivalent $U(n)$~bundles over $\mathcal M$~the expression
\begin{align*}
\text{Vect}_n(\mathcal M, \mathbb C)=\pi\left[\mathcal M,G_{n}(\mathcal C^\infty)\right],
\end{align*}
which is known for some rather simple base manifolds. In particular for spheres $S^d$, there is a trick to calculate $\text{Vect}_n(S^d,\mathbb C)$: $S^d$~can always be decomposed into two hemispheres which are individually trivial. The homotopy of a bundle over $S^d$~is thus determined by the clutching function $f_c$~defined in the overlap $S^{d-1}$~of the two hemispheres, i.e., along the equator of $S^d$. Physically, $f_c$~translates a local gauge choice on the upper hemisphere into a local gauge choice on the lower hemisphere and is thus a function $f_c:S^{d-1}\rightarrow U(n)$. The group of homotopy classes of such functions is by definition $\pi_{d-1}(U(n))$. Interestingly, for $n>\frac{d-1}{2}$, these groups are given by
\begin{align}
\text{Vect}_n(S^d, \mathbb C)=\pi_{d-1}(U(n))=\left\{ \begin{array}{l}{\mathbb Z,~d-1~\text{odd}}\\ {\left\{0\right\},~d-1~\text{even}}\end{array}\right.
\label{eqn:complexbott}
\end{align}
This periodicity of two in $(d-1)$~is known as the complex Bott periodicity. The physical meaning of Eq. (\ref{eqn:complexbott}) is the following: In the unitary universality class A, there is an integer topological invariant in even spatial dimension (e.g., QAH in $d=2$)~and no TSM in odd spatial dimension.\\

This classification has two shortcomings. First, it cannot be readily generalized to other CAZ classes at this simple level. Second, only systems with the same number of occupied bands $n$~can be compared. However, adding some topologically trivial bands to the system should yield a system in the same equivalence class, if those bands can be considered as inert, i.e., if they do not change the low energy physics close to the Fermi surface. Both shortcomings can be overcome in the framework of K-Theory \cite{Karoubi,Nash}.

\subsection{K-Theory approach to a complete classification}
K-Theory \cite{Karoubi,Nash} is concerned with vector bundles which have a ``sufficiently large'' fiber dimension. This means, that topological defects which can be unwound by just increasing the fiber dimension are not visible in the resulting classification scheme. This is physically reasonable, as trivial occupied bands from inner localized shells for example increase the number of bands as compared to the effective low energy models under investigation. Models of different number of such trivial bands should be comparable in a robust classification scheme. In Ref. \onlinecite{HorovaKTheory}, K-Theory was used to discuss analogies between the emergence of D-branes in superstring theory and the stability of Fermi-surfaces in non-relativistic systems. The use of K-Theory for the classification of TSM has been pioneered in Refs. \onlinecite{KaneMele2005a,KaneMele2005b} for the QSH state and more systematically been discussed for general TSM in Ref. \onlinecite{KitaevPeriodic}. This analysis in turn can be understood as a special case of a twisted equivariant K-Theory as has been reported very recently \cite{FreedMoore}.\\

\subsubsection*{Crash-course in K-Theory}
The direct sum of two vector bundles $E\oplus F$~is the direct sum of their fibers over each point. This addition has only a semi-group structure, since $E\oplus G=F\oplus G\nRightarrow E\simeq F$. A minimal counterexample is given by $E=TS^2,~F=S^2\times \mathbb R^2$. $F$~is clearly trivial, whereas $TS^2$, the tangent bundle of $S^2$, i.e., the disjoint union of all tangent planes of $S^2$, is well known to be non-trivial. However, adding $NS^2$, the bundle of normal vectors to $S^2$, to both bundles $E,F$~we obtain the same trivial bundle $S^2\times \mathbb R^3$. This motivates the concept of stable equivalence
\begin{align}
E\overset{s}{\simeq}F\Leftrightarrow E\oplus Z^m\simeq F\oplus Z^n,
\label{eqn:defstable}
\end{align}
where $Z^n=\mathcal M\times K^n,~K=\mathbb R,\mathbb C$~is the trivial bundle over the fixed base manifold $\mathcal M$, which plays the role of an additive zero as far as stable equivalence is concerned. We denote the set of $K$-vector bundles over $\mathcal M$~by $\mathcal V_K(\mathcal M)$~in the following. Note that stably equivalent bundles can have different fiber dimension, as $m\ne n$~in general in Eq. (\ref{eqn:defstable}). The benefit of this construction is:
\begin{align}
E\oplus G=F\oplus G\Rightarrow E\overset{s}{\simeq}F
\label{eqn:stableeq}
\end{align}
This is because for vector bundles on a smooth manifold every bundle can be augmented to a trivial bundle, i.e.,
\begin{align}
\forall_G\exists_{H,l}~G\oplus H=Z^l.
\label{eqn:stableinv}
\end{align}
Eq. (\ref{eqn:stableeq}) naturally leads to the notion of a subtraction on $\mathcal V_K(\mathcal M)$~by virtue of the Grothendieck construction: Consider the pairs $(E_1,E_2)\in \mathcal V_K(\mathcal M)\times\mathcal V_K(\mathcal M)$~and define the equivalence relation
\begin{align}
 (E_1,E_2)\sim(F_1,F_2)\Leftrightarrow \exists H~F_1\oplus E_2\oplus H \simeq  E_1\oplus F_2\oplus H.
 \label{eqn:Grothendieck}
 \end{align}
Looking at Eq. (\ref{eqn:Grothendieck}), we can intuitively think of the equivalence class $(E_1,E_2)_\sim$~as the formal difference $E_1-E_2$.  We now define the K-group as the quotient
\begin{align}
K(\mathcal M)=\left(\mathcal V_K(\mathcal M)\times\mathcal V_K(\mathcal M)\right)/\sim,
\end{align}
 which identifies all formal differences that are equivalent in the sense of Eq. (\ref{eqn:Grothendieck}).  Due to Eq. (\ref{eqn:stableinv}), every group element in $K(\mathcal M)$~can be represented in the form $(E,Z^n)$. However, $(E,Z^n)\nsim (E,Z^m)$~for $n\ne m$. We define the virtual dimension of $(E,F)$~as $d_v=\text{rk}(E)-\text{rk}(F)$,~where rk denotes the rank, i.e., the fiber dimension of a vector bundle. By restricting $K(\mathcal M)$~to elements with $d_v=0$, we obtain the restricted K-group $\tilde K(\mathcal M)=\left\{g\in K(\mathcal M)|d_v(g)=0\right\}$. $\tilde K(\mathcal M)$~is isomorphic to the set of stable equivalence classes of $\mathcal V_K(\mathcal M)$. Up to now, the construction has been independent of the field over which the vector spaces are defined. In the following, we will distinguish the real and complex K-groups $K_{\mathbb R}(\mathcal M), K_{\mathbb C}(\mathcal M)$. Physically, $K_{\mathbb C}$~will be employed to characterize systems without anti-unitary symmetries, whereas $K_{\mathbb R}$~is relevant for systems in which at least one anti-unitary symmetry imposes a reality constraint on the $k$-space.\\
 
 A crucial notion in K-Theory which is also our main physical motivation to study it is that of the stable range. The idea is that at sufficiently large fiber dimension $n$ no ``new'' bundles can be discovered by looking at even larger fiber dimension. Sufficiently large in terms of the dimension $d$~of $\mathcal M$~means $n\ge n_{\mathbb C}=d/2+1$~for the complex case and $n\ge n_{\mathbb R}=d+1$~for the real case, respectively. More formally, every bundle $E$~with $n>n_K$~can be expressed as a sum 
 \begin{align}
 E\simeq F\oplus Z^{n-n_K}
 \label{eqn:stablerange}
 \end{align}
 of a bundle $F$~with fiber dimension $n_K$~and a trivial bundle for $K=\mathbb R,\mathbb C$. Since clearly $E\overset{s}{\simeq}F$~(see Eq. (\ref{eqn:stableeq})), this means that all stable equivalence classes have representatives in fiber dimension $n\le n_K$. Furthermore, a situation like our counterexample above where we augmented two non-isomorphic bundles by the same trivial bundle $NS^2$~to obtain the same trivial bundle cannot occur in the stable range. That is to say $F$~as appearing in Eq. (\ref{eqn:stablerange}) is uniquely defined up to isomorphisms. The stable range hence justifies the approach of $K$-theory of ignoring fiber dimension when defining the stable equivalence. The key result in the stable range which connects K-Theory to our goal of classifying all inequivalent vector bundles with sufficiently large but arbitrary fiber dimension on equal footing reads \cite{Nash}
 \begin{align}
 \tilde K_K(\mathcal M)=\text{Vect}_n(\mathcal M,K) = \pi\left[\mathcal M,G_{n}(K^\infty)\right]\quad \forall_{n\ge n_K}.
 \label{eqn:classnind}
 \end{align}
The complex Bott periodicity Eq. (\ref{eqn:complexbott}) with period $p_{\mathbb C}=2$~has a real analogue concerning the homotopy groups of $\text{O}(n)$~with period $p_{\mathbb R}=8$. This immediately implies in the language of K-Theory
\begin{align}
\tilde K_K(S^{d+p_K})=\tilde K_K(S^d),~K=\mathbb R,\mathbb C.
\label{eqn:trivKbott}
\end{align}
We define 
\begin{align}
\tilde K_K^{-d}(\mathcal M)= \tilde K_K(S^d \mathcal M),
\label{eqn:kn}
\end{align}
where $S$~is the reduced suspension (see Refs. \onlinecite{Nash,NashSen} for a detailed discussion) which for a sphere $S^k$ indeed satisfies $SS^k=S^{k+1}$. The stronger version of the Bott periodicity in K-Theory now reads \cite{Nash}
\begin{align}
\tilde K_K^{-d-p_K}(\mathcal M)=\tilde K_K^{-d}(\mathcal M),~K=\mathbb R,\mathbb C,
\label{eqn:Kbott}
\end{align}
which only for $\mathcal M=S^l$~trivially follows from Eq. (\ref{eqn:trivKbott}). Using this periodicity, the definition of $K_K^{-d}$~in Eq. (\ref{eqn:kn}) can be formally extended to $d\in \mathbb Z$. 

\begin{table}[ht]
\centering
\begin{tabular}{|l|l|}\hline
 Class&Classifying Space  \\ \hline
 A &$C_0=U(n+m)/\left(U(n)\times U(m)\right)$\\
 AIII &$C_1=U(n)$\\ \hline
 AI &$R_0=O(n+m)/\left(O(n)\times O(m)\right)$\\
 BDI&$R_1=O(n)$\\
 D&$R_2=U(2n)/U(n)$\\
 DIII&$R_3=U(2n)/Sp(2n)$\\
 AII &$R_4=Sp(n+m)/\left(Sp(n)\times Sp(m)\right)$\\ 
 CII&$R_5=Sp(n)$\\ 
 C&$R_6=Sp(2n)/U(n)$\\
 CI&$R_7=U(n)/O(n)$\\ \hline
\end{tabular}
\caption{\label{tab:classspaces} Table of all classifying spaces $C_q,~R_q$~of complex and real K-Theory, respectively. The first column denotes the CAZ symmetry class. From top to bottom, the next complex/real classifying space is the loop space of its predecessor, i.e., $C_{q+1}=\Omega C_q (\text{mod }2)$, $R_{q+1}=\Omega R_q (\text{mod }8)$
}
\end{table}

\subsubsection*{The Bott clock}
From the very basic construction of homotopy groups the following identities for the homotopy of a topological space $X$~are evident:
\begin{align*}
\pi_d(X)=\pi\left[S^d,X\right]=\pi\left[S S^{d-1},X\right]=\pi\left[S^{d-1},\Omega X\right],
\end{align*}
where $\Omega X$~denotes the loop space \cite{Nash} of $X$, i.e., the space of maps from $S^1$~to $X$. Iterating this identity gives $\pi_d(X)=\pi_0(\Omega^nX)$~using the complex Bott periodicity (\ref{eqn:complexbott}), we immediately see that counting the connected components $\pi_0\left(U(n)\right),\pi_1\left(U(n)\right)=\pi_0\left(\Omega U(n)\right)$~of the unitary group and its first loop space, we can classify all $U(n)$~vector bundles over $S^d$~in the stable range, i.e., with $n>\frac{d}{2}$. The real analogue of the Bott periodicity with period $p_{\mathbb R}=8$~leads to analogous statements for $O(n)$~bundles over $S^d$~which depend only on the connected components of $O(n)$~and its first seven loop spaces $\Omega^i O(n),~i=1,\ldots,7$ (see Tab. \ref{tab:classspaces}). This defines a Bott clock with two ticks for the complex case and eight ticks for the real case, respectively. Interestingly, these ten spaces, for the complex and real case together, are precisely the ten Cartan symmetric spaces in 
which the time evolution operators associated with Hamiltonians in the ten CAZ classes lie. After this 
observation, only two points are 
missing until a complete classification of all TSM of continuum models can be achieved. The first point is a subtlety related to the interdependence of the two wave vectors $k$~and $-k$~as shown in Eq. (\ref{eqn:symBloch}), which makes the real Bott clock tick counter clockwise. The second point is the inclusion of symmetry constraints into the scheme which leads to the clockwise ticking Clifford clock (see Eq. (\ref{eqn:ciffclock}) below). The combination of both implies that the topological invariant of a continuum model of dimension $d$~in the CAZ class $q$~only depends on the difference $q-d~(\text{mod}~8)$~for the eight real classes and on $q-d~(\text{mod}~2)$~for the two complex classes A and AIII, respectively.   
\subsubsection*{Reality and $\mathbf{k}$-space topology}
For systems which obey anti-unitary symmetries the real structure of the Hamiltonian $H(x)$~is most conveniently accounted for in its Majorana representation. $H(x)$~can in this representation be expressed in terms of a real antisymmetric $2n\times 2n$-matrix~$B$,
\begin{align}
\Psi^\dag(x)H(x)\Psi(x)=\frac{i}{4}B^{ij}c_{x,i}c_{x,j},
\label{eqn:majrep}
\end{align}
where $c_{x,i},~i=1,\ldots,2n$~are the Majorana operators representing the $n$~fermion species at $x$. On Fourier transform, $\mathcal H = \int\Psi^\dag H \Psi$~can be written as \cite{KitaevPeriodic}
\begin{align}
\mathcal H =\frac{i}{4}\int \text{d}^dk A^{ij}(k)c_{-k,i}c_{k,j},
\label{eqn:majk}
\end{align}
where $A$~is skew hermitian and satisfies
\begin{align}
A^*(k)=A(-k).
\label{eqn:Arealstruc}
\end{align}
Eq. (\ref{eqn:Arealstruc}) naturally leads to a real vector bundle structure as defined in Ref. \onlinecite{AtiyahReal} for the bundle of eligible $A$-matrices over the $k$-space $(\mathbb R^d,\tau)$, where the involution $\tau$~(see Ref. \onlinecite{AtiyahReal}) is given by $k\mapsto -k$. On one-point compactification, this real $k$-space becomes a sphere $\bar S^d=(S^d,\tau)$~with the same involution \cite{KitaevPeriodic}. Whereas the ordinary sphere $S^d$~can be viewed as a reduced suspension $S$~of $S^{d-1}$~over the real axis, $\bar S^d$~can be understood as the reduced suspension $\bar S$~of $\bar S^{d-1}$~over the imaginary axis. This picture is algebraically motivated by comparing the involution $\tau$~to the ordinary complex conjugation which, restricted to the imaginary axis of the complex plane is of the same form. Interestingly, in the language of definition (\ref{eqn:kn}), $\bar S$~plays the role of an inverse to $S$~\cite{AtiyahReal,KitaevPeriodic}, i.e.,
\begin{align*}
\tilde K_{\mathbb R}(\mathcal M)= \tilde K_{\mathbb R}^{-1}(\bar S \mathcal M)=\tilde K_{\mathbb R}(S\bar S\mathcal M).
\end{align*}
This means that the Bott clock over $\bar S^d$~is reversed as compared to its analogue over $S^d$.
\subsubsection*{Real K-theory and the Clifford clock}
The main reason for the real construction of Eq. (\ref{eqn:majk}) is that the anti-unitary symmetry constraints yielding the eight real CAZ classes (all except A and AIII) can be distinguished in terms of anti-commutation relations of the $A$-matrix with real Clifford generators \cite{Karoubi,KitaevPeriodic}. At a purely algebraic level, these constraints can be transformed so as to be expressed only in terms of positive Clifford generators \cite{KitaevPeriodic}, i.e., generators that square to plus identity. We call the restricted K-group of a vector bundle of $A$-matrices over $\mathcal M$~that anti-commute with $q$~positive Clifford generators $\hat K^q_{\mathbb R}(\mathcal M)$. Interestingly \cite{Karoubi,KitaevPeriodic},
\begin{align}
\hat K^q_{\mathbb R}(\mathcal M)\simeq \tilde K^{-q}_{\mathbb R}(\mathcal M).
\label{eqn:ciffclock}
\end{align}
Eq. (\ref{eqn:ciffclock}) defines a Clifford clock that runs in the opposite direction as the $\bar S^d$~Bott clock. This algebraic phenomenon explains the full periodic structure of the table of TSM of continuum models (see Tab. \ref{tab:classclean}). The classifying spaces of $A$-matrices for systems that anti-commute with $q$~Clifford generators are shown in Tab. \ref{tab:classspaces}. 
\subsubsection*{Periodic systems}
The classification of periodic systems is much more complicated from a mathematical point of view. Their base space is the real Brillouin zone $\bar T^d=(T^d,\tau)$, where the involution $\tau$~giving rise to the real structure is again given by $k\mapsto -k$. For $\bar T^d$~the reduced suspension does not provide a trivial relation between the K-Theory of different spatial dimension like $\bar S \bar S^d= \bar S^{d+1}$~for the base space of continuum models. The general calculation of all relevant K-groups over $\bar T^d$~has been reported in Ref. \onlinecite{KitaevPeriodic}. Interestingly, the resulting groups always contain the respective classification of continuum models in the same symmetry class as an additive component. Additionally, the topological invariants of weak TSM, i.e., TSM which are only present in translation-invariant systems, can be inferred. The Clifford clock defined in Eq. (\ref{eqn:ciffclock}) is independent of the base space and hence still applicable. Here, we only review the general 
result 
calculated in Ref. \onlinecite{KitaevPeriodic}
\begin{align}
K^{-q}_{\mathbb R}(\bar T^d)\simeq K^{-q}_{\mathbb R}(\bar S^d)\oplus\left(\bigoplus\limits_{s=0}^{d-1}\begin{pmatrix}d\\s\end{pmatrix}K^{-q}_{\mathbb R}(\bar S^s)\right).
\label{eqn:periodicTSM}
\end{align}
The second term on the right hand side of Eq. (\ref{eqn:periodicTSM}) entails the notion of so called weak topological insulators which are obviously due to TSM in lower dimensions. To name the most prominent example, the $\mathbb Z_2$~invariant characterizing the QSH insulator in $d=2$~in the presence of TRS, CAZ class AII, yields a $3\mathbb Z_2$~topological invariant  characterizing the weak topological insulators with the same symmetry in $d=3$.
\subsubsection*{Lattice systems with disorder}
In a continuum model, disorder that is not too short ranged so as to keep the $k$-space compactification for large $k$~valid, can be included into the model system without changing the classification scheme. However, perturbing a translation-invariant lattice system with disorder also gives its $k$-space (now defined in terms of a discrete Fourier transform) a discrete lattice structure which is not directly amenable to investigation in the framework of K-Theory which we only defined over smooth base-manifolds. Ref. \onlinecite{KitaevPeriodic} shows that a Hamiltonian featuring localized states in the energy gap can be transformed into a gapped Hamiltonian upon renormalization of parameters. The physical consequence of this statement is that only a mobility gap is needed for the classification of a TSM and no energy gap in the density of states. Furthermore, Ref. \onlinecite{KitaevPeriodic} argues without explicit proof that the classification problem of gapped lattice systems without translation invariance is equivalent to the classification problem of continuum models. 
This 
statement agrees with the physical intuition that the breaking of translation-invariance must remove the additional structure of weak TSM as described for periodic systems by Eq. (\ref{eqn:periodicTSM}).     

\section{Calculation of topological invariants of individual systems}
\label{sec:topcalc}
In Section \ref{sec:symmclass}, we have shown how many different TSM can be expected in a given spatial dimension and CAZ class. Now, we outline how insulating systems within the same CAZ class and dimension can be assigned a topological equivalence class in terms of their adiabatic connection defined in Eq. (\ref{eqn:Katoconnection}) and their adiabatic curvature defined in Eq. (\ref{eqn:Katocurv}), respectively. A complete case by case study in terms of Dirac Hamiltonian representatives of all universality classes of this problem has been reported in Ref. \onlinecite{RyuLudwig}. We outline the general patterns relating the classification of neighboring (see Tab. \ref{tab:classclean}) universality classes following the analysis in Refs. \onlinecite{QiTFT,Schnyder2008,RyuLudwig}. Interestingly, all topological invariants can be calculated using only complex invariants, namely Chern numbers and chiral unitary winding numbers. The anti-unitary symmetries are accounted for by the construction of a 
dimensional hierarchy in Section \ref{sec:dimhier} starting from a so called parent state in each symmetry class for which the complex classification concurs with the real classification. In Section \ref{sec:tbc}, we show how the topological invariants can be defined for disordered systems with the help of twisted boundary conditions. Furthermore, we discuss a generalization of the non-interacting topological invariants to interacting systems in Section \ref{sec:int}.

\subsection{Systems without anti-unitary symmetries}
\label{sec:complexclasses}
\subsubsection*{Chern numbers of unitary vector bundles}
Eq. (\ref{eqn:unipullback}) shows that every $U(n)$~bundle $E\rightarrow \mathcal M$~can be represented as a pullback from the universal bundle $\xi\rightarrow G_n(\mathbb C^\infty)$~by some bundle map $\hat f$. Chern classes are de Rham cohomology classes, i.e., topological invariants \cite{BottTu} that are defined as the pullback of certain cohomology classes of the classifying space $G_n(\mathbb C^\infty)$. The cohomology ring $H^*\left(G_n(\mathbb C^\infty)\right)$~consists only of even classes and is generated by the single generator $\tilde c_j\in H^{2j}\left(G_n(\mathbb C^\infty)\right),~j=1,\ldots,n$~for every even cohomology group \cite{Sato}. The Chern classes  $c_i$~ of $E$~are defined as the pullback $c_i = f^* \tilde c_i$~from the classifying space by the map $f:\mathcal M\rightarrow G_n(\mathbb C^\infty)$~associated with the bundle map $\hat f$. Due to the Chern-Weyl theorem \cite{Nakahara}, Chern classes can be expressed in terms of the curvature, i.e., in our case, the adiabatic curvature 
$\mathcal 
F$~defined in Eq. (\ref{eqn:Katocurv}) of $E$. Explicitly, the total Chern class $c$~can be expressed as \cite{Nakahara}
\begin{align}
c=\det\left(1+\frac{i\mathcal F}{2\pi}\right)=1+c_1(\mathcal F)+c_2(\mathcal F)\ldots .
\label{eqn:totalchern}
\end{align}
The determinant is evaluated in gauge space and products of $\mathcal F$~are understood to be wedge products. $c_j$~is the monomial of order $j$~in $\mathcal F$. Obviously, $c_j$~is a $2j$-form and can only be non-vanishing for $2j\le d$, where $d$~is the dimension of the base manifold $\mathcal M$, i.e., the spatial dimension of the physical system. Another characteristic class which generates all Chern classes is the Chern character \cite{Nakahara}
\begin{align}
\text{ch} = \text{Tr}\left[\text{e}^{\frac{i\mathcal F}{2\pi}}\right] =1+ \text{ch}_1(\mathcal F)+\text{ch}_2(\mathcal F)+\ldots .
\label{eqn:chernchar}
\end{align}
Due to their importance for later calculations, we explicitly spell out the first two Chern characters $\text{ch}_1=\text{Tr}\left[\frac{i\mathcal F}{2\pi}\right],~\text{ch}_2=-\frac{1}{8\pi^2}\text{Tr}\left[\mathcal F \wedge \mathcal F\right]$. Importantly, for even $d=2p$, the integral
\begin{align*}
\mathcal C_p= \int_\mathcal M\text{ch}_p
\end{align*}
yields an integer, the so called $p$-th Chern number \cite{ChoquetBruhat}. These Chern numbers characterize systems in the unitary symmetry class A which can only be non-trivial in even spatial dimension (see Tab. \ref{tab:classclean}).

\subsubsection*{Winding numbers of chiral unitary vector bundles}
In Section \ref{sec:symmclass}, we have shown that the classifying space for a chiral unitary (AIII) system is given by $U(n)$~and that the topological sectors are defined by homotopically distinct maps $k\mapsto q(k)\in U(n)$. Now, we discuss how to assign an equivalence class to a given  map $q$~by calculating its winding number \cite{Redlich1984,VolovikQH3HE,Golterman1993} following Ref. \onlinecite{RyuLudwig}. From Tab. \ref{tab:classclean} it is clear that only in odd spatial dimension $d=2j-1$~there can be a non-trivial winding number. We define
\begin{align}
w_{2j-1}^q= \frac{(-(j-1)!)}{(2j-1)!(2\pi i)^{j}}\text{Tr}\left[(q^{-1}dq)^{2j-1}\right],
\label{eqn:winddens}
\end{align}
which has been dubbed winding number density \cite{RyuLudwig}. Integrating this density over the odd-dimensional base manifold $\mathcal M$~representing the $k$-space of the physical system, we get the integral winding number $\nu_{2j-1}$
\begin{align}
\nu_{2j-1}=\int_{\mathcal M}w_{2j-1}^q,
\label{eqn:windingdef}
\end{align}
which is well known to measure the homotopy of the map $k\mapsto q(k)$.

\subsubsection*{Relation between chiral winding number and Chern Simons form}
So far, the relation between the adiabatic connection of a chiral system and its topological invariant has not been made explicit. Since characteristic classes like Chern characters are closed $2j$-forms, they can locally be expressed as exterior derivatives of $(2j-1)$-forms. These odd forms are called the Chern Simons forms associated with the even characteristic class \cite{ChernSimons1974,Nakahara}. For the $j$-th Chern character $\text{ch}_j$, which is a $2j$~form, the associated Chern Simons form $\mathcal Q_{2j-1}$~reads \cite{Nakahara}
\begin{align}
\mathcal Q_{2j-1}(\mathcal A,\mathcal F_t)=\frac{1}{(j-1)!}\left(\frac{i}{2\pi}\right)^{j}\int_0^1\text{dt}~\text{STr}\left[\mathcal A,\mathcal F_t^{j-1}\right],
\label{eqn:csdef}
\end{align}
where $\mathcal F_t=t\mathcal F +(t^2-t)\mathcal A\wedge \mathcal A$~is the curvature of the interpolation $t\mathcal A$~between the zero connection and $\mathcal A$~and STr denotes the symmetrized trace. Explicitly, we have $\mathcal Q_1=\frac{i}{2\pi}\text{Tr}\left[\mathcal A\right],~\mathcal Q_3=-\frac{1}{8\pi^2}\text{Tr}\left[\mathcal Ad\mathcal A +\frac{2}{3}\mathcal A^3\right]$.\\

 It is straightforward to show \cite{RyuLudwig}, that in a suitable gauge, the Berry connection of a chiral bundle yields $\mathcal A^B=\frac{1}{2}qdq^\dag$, where $q\in U(n)$~is again the chiral map characterizing the system. This is not a pure gauge due to the factor $\frac{1}{2}$~which entails that the associated curvature $\mathcal F^B$ does not vanish. Plugging $\mathcal A^B$~and $\mathcal F^B$~into Eq. (\ref{eqn:csdef}) immediately yields \cite{RyuLudwig}
 \begin{align}
 \mathcal Q_{2j-1}(\mathcal A^B,\mathcal F_t^B)=\frac{1}{2}w_{2j-1}^q.
 \label{eqn:relcswn}
 \end{align}
Eq. (\ref{eqn:relcswn}) directly relates the winding number density to the Chern Simons form. We define the Chern Simons invariant of an odd dimensional system as
\begin{align*}
\text{CS}_{2j-1}= \int_{\mathcal M}\mathcal Q_{2j-1}~(\text{mod }1),
\end{align*}
where (mod $1$)~accounts for the fact that $\int_{\mathcal M}\mathcal Q_{2j-1}$~has an integer gauge dependence due to $\pi_{2j-1}\left(U(n)\right)=\mathbb Z$~for $n>j$. Looking back at Eq. (\ref{eqn:windingdef}), we immediately get
\begin{align*}
\nu_{2j-1} (\text{mod }2)=2\text{CS}_{2j-1} (\text{mod }2).
\end{align*}
We note that the (mod $2$) can be dropped if we fix the gauge as described above to $\mathcal A^B=\frac{1}{2}qdq^{-1}$. This establishes the desired relation between the winding number of a chiral unitary system and its adiabatic curvature. 

\subsection{Dimensional hierarchy and real symmetry classes}
\label{sec:dimhier}
Until now, we have only discussed how to calculate topological invariants of systems in the complex symmetry classes A and AIII. Interestingly, for some real universality classes, the classification in the presence of anti-unitary symmetries concurs with the unitary classification (see Tab. \ref{tab:hierarchy}). The first known example of this type is in the symplectic class AII in $d=4$~which is characterized by the second Chern number of the corresponding complex bundle \cite{Avron1988,ZhangHu4DQH}. Another example of this kind is the $p+ip$~superconductor in $d=2$~and symmetry class D which is characterized by its first Chern number, i.e., in the same way as the QAH effect in class A. In odd dimensions similar examples exist for real chiral classes, e.g., for DIII in $d=3$, where the winding number is calculated using Eq. (\ref{eqn:winddens}) in the same way as for the chiral unitary class AIII in the same dimension. All the topological invariants just mentioned are integer invariants. In some 
universality 
classes, these integers can 
only assume even values (see Tab. \ref{tab:hierarchy}).
 For 
physically relevant dimensions, i.e., $d=1,2,3$, these exceptions are CII in $d=1$, C in $d=2$, and CI in $d=3$. All other states where the complex and the real classification concur, can be viewed as parent states of a dimensional hierarchy within the same symmetry class from which all $\mathbb Z_2$~invariants appearing in Tab. \ref{tab:hierarchy} can be obtained by dimensional reduction. This approach was pioneered in the seminal work by Qi, Hughes, and Zhang \cite{QiTFT}.\\
\begin{table*}[ht]
\centering
\begin{tabular}{|l|cccccccc|}\hline
 Class&$d=1$&$d=2$&$d=3$&$d=4$&$d=5$&$d=6$&$d=7$&$d=8$\\ \hline
 A &0&$\mathbb Z$&0&$\mathbb Z$&0&$\mathbb Z$&0&$\mathbb Z$\\
 \bf{AIII} &$\mathbb Z$&0&$\mathbb Z$&0&$\mathbb Z$&0&$\mathbb Z$&0\\ \hline
 AI&0&0&0&$2\mathbb Z$&0&$\mathbb Z_2$&$\mathbb Z_2$&$\boxed{\mathbb Z}$\\
 \bf{BDI}&$\boxed{\mathbb Z}$&0&0&0&$2\mathbb Z$&0&$\mathbb Z_2$&$\mathbb Z_2$\\
 D&$\mathbb Z_2$&$\boxed{\mathbb Z}$&0&0&0&$2\mathbb Z$&0&$\mathbb Z_2$\\
 \bf{DIII}&$\mathbb Z_2$&$\mathbb Z_2$&$\boxed{\mathbb Z}$&0&0&0&$2\mathbb Z$&0\\
 AII&0&$\mathbb Z_2$&$\mathbb Z_2$&$\boxed{\mathbb Z}$&0&0&0&$2\mathbb Z$\\ 
 \bf{CII}&$2\mathbb Z$&0&$\mathbb Z_2$&$\mathbb Z_2$&$\boxed{\mathbb Z}$&0&0&0\\ 
 C&0&$2\mathbb Z$&0&$\mathbb Z_2$&$\mathbb Z_2$&$\boxed{\mathbb Z}$&0&0\\
 \bf{CI}&0&0&$2\mathbb Z$&0&$\mathbb Z_2$&$\mathbb Z_2$&$\boxed{\mathbb Z}$&0\\ \hline
\end{tabular}
\caption{\label{tab:hierarchy} Table of all groups of topological equivalence classes. The first column denotes the symmetry class, divided into two complex classes without any anti-unitary symmetry (top) and eight real classes with at least one anti-unitary symmetry (bottom). Chiral classes are denoted by bold letters. The parent states of dimensional hierarchies are boxed. For all non-chiral boxed states, the classification concurs with that of class A in the same dimension. For all chiral boxed states, the classification concurs with that of class AIII in the same dimension. $2\mathbb Z$~indicates that the topological integer can only assume even values in some cases. Such states are never parent states.   
}
\end{table*}
\begin{figure}
	\centering
	\includegraphics[width=0.35\textwidth]{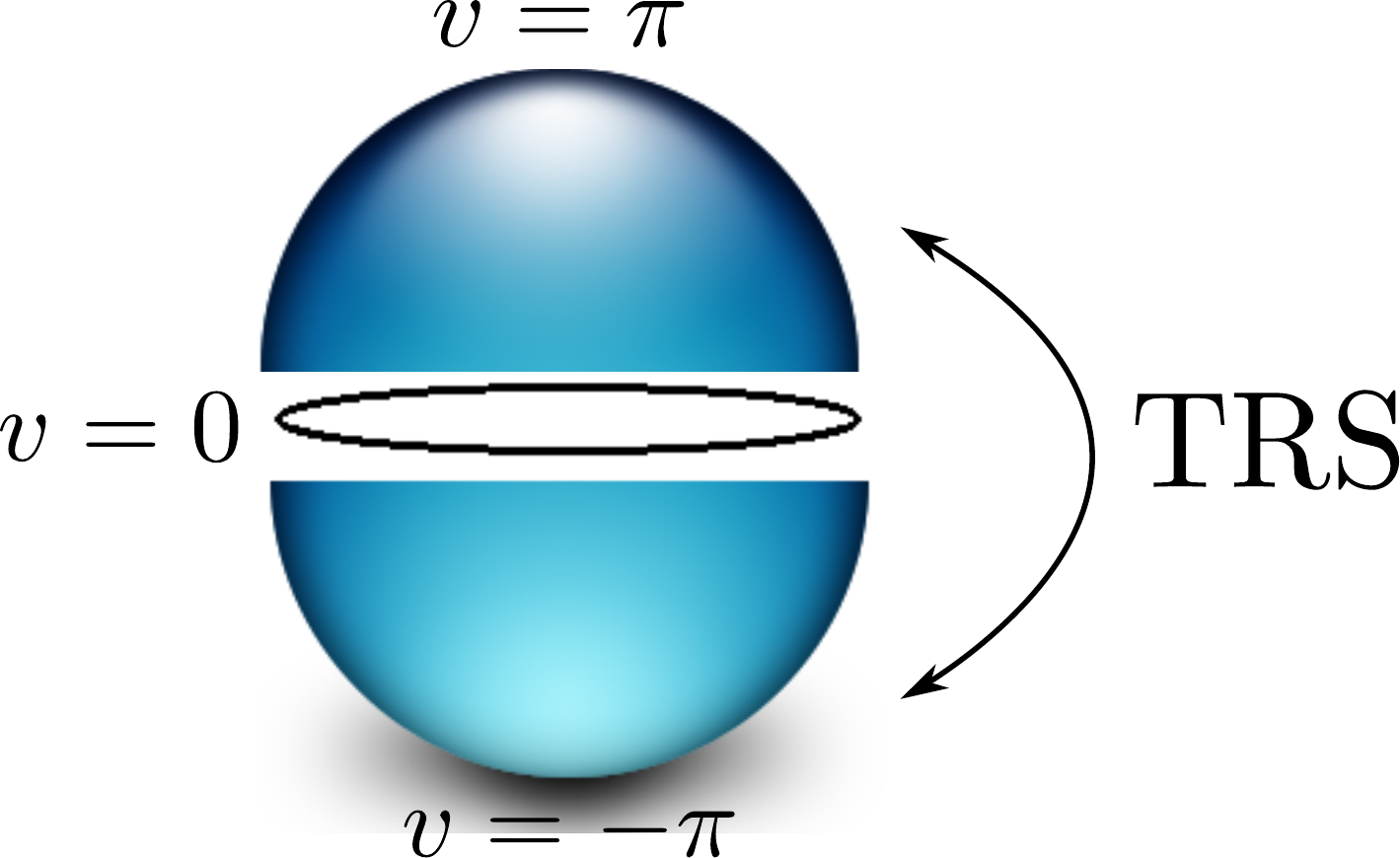}
	\caption{Illustration of the WZW dimensional extension. The circle at $v=0$~represents the physical system. The poles at $v=\pm \pi$~represent the trivial reference system without $k$-dependence. The two interpolations are conjugated by an anti-unitary symmetry, here exemplary denoted by TRS.}
	\label{fig:wzw}
\end{figure}

The general idea is more intuitive if we consider the parent state as a Wess-Zumino-Witten (WZW) dimensional extension \cite{WZW,TopologicalOrderParameter} of the lower dimensional descendants instead of thinking of a dimensional reduction from the $d$-dimensional parent state to its descendants. This works as follows: We fix a localized $(d-1)$-dimensional insulator without any hopping that satisfies the required anti-unitary symmetries as a trivial reference state. This reference state is described by the $k$-independent Bloch Hamiltonian $h_0$. The $(d-1)$-dimensional physical system of interest is characterized by the Bloch Hamiltonian $h(k)$. Then, we interpolate by varying the parameter $v$~between the $(d-1)$-dimensional physical system ($v=0$)~and the trivial state ($v=\pi$)~without closing the insulating gap. However, the intermediate $(d-1)$-dimensional system at fixed $v\ne 0,\pi$~might well break the required anti-unitary symmetries. The crucial step is now to do the interpolation for $v\in \left[
0,\pi\right]$~and $v\in \left[-\pi,0\right]$~in a symmetry conjugated way (see Fig. \ref{fig:wzw}). That is to say, we require our $(d-1)$-dimensional system of interest and the resulting $d$-dimensional extended system to be in the same CAZ class. This $d$-dimensional system is characterized by the Bloch Hamiltonian $h(k,v)$. The $v\in \left[-\pi,0\right]$~and the $v\in \left[0,\pi\right]$~half of the extended $k$-space then are not independent of each other but give equal contributions to the integer topological invariant of the $d$-dimensional extended system \cite{QiTFT,RyuLudwig}. One might now ask to which extend the resulting integer invariant of the extended system depends on our choice of the interpolation $h(k,v)$~between $h(k)=h(k,v=0)$~and $h_0=h(k,v=\pm \pi)$. To answer this question, one considers two interpolations $h(k,v),\tilde h(k,v)$. It is then elementary to show \cite{QiTFT} that the difference between the integer invariants of these two $d$-dimensional systems is an even integer. This 
implies that a $\mathbb Z_2$~information, namely the parity of the integer invariant associated with the extended system, is well defined only in terms of the physical system with spatial dimension $(d-1)$.\\

A similar procedure can be repeated a second time to obtain a $\mathbb Z_2$~invariant for a $(d-2)$~dimensional second descendant \cite{QiTFT}. From the procedure just sketched, it is obvious why the exceptional phases which are characterized by an even integer are not parent states of such a dimensional hierarchy. The generic constructions for all possible classes can be found in Refs. \onlinecite{QiTFT,RyuLudwig}. With that, we are provided with a general and fairly explicit 
recipe for 
the practical calculation of the topological invariants for all possible CAZ classes in all spatial dimensions.

\subsection{Bulk invariants of disordered systems and twisted boundary conditions}
\label{sec:tbc}
Our practical calculation of topological invariants so far has been focused on periodic systems with a BZ, i.e., $\mathcal M=T^d$~and continuum models where the $k$-space can be compactified to a sphere, i.e.,  $\mathcal M =S^d$. As already pointed out in Section \ref{sec:symmclass}, the situation is more complicated for disordered lattice models. Seminal progress along these lines was reported for the quantum Hall state by Niu {\emph et al.} in 1985 \cite{Niu1985}. These authors use twisted boundary conditions (TBC) to define the quantized Hall conductivity $\sigma_{xy}$~for a 2D system as a topological invariant only requiring a bulk mobility gap. We briefly review their analysis and propose the framework of TBC as a general recipe to calculate topological invariants for disordered systems.\\

The Hall conductivity $\sigma_{xy}$~resulting from a linear response calculation at zero temperature yields
\begin{align*}
 \sigma_{xy}=\frac{-2}{A}\textrm{Im}\sum_{n\ne 0}\frac{\langle 0\rvert \mathcal H_x\lvert n\rangle\langle n\rvert\mathcal H_y\lvert 0\rangle}{(E_n-E_0)^2},
\end{align*}
where $\lvert 0\rangle$~is the many body ground state, $A$~is the area of the system, and $\mathcal H_i=\frac{\partial \mathcal H}{\partial k^i}$. In the presence of a magnetic field, translation-invariance is defined in terms of the magnetic translation operator $T_B$~\cite{Kohmoto1985} which concurs with the ordinary translation operator $T(a)=\text{e}^{ia\hat k}$~in the absence of a magnetic field. TBC now simply mean that a (magnetic) translation by the system length $L_j$~in $j$-direction gives an additional phase factor $\text{e}^{i\phi_j}$. $\phi_j$~is called the twisting angle in $j$-direction. Gauging away this additional phase to obtain a wave function with periodic boundaries amounts to a gauge transformation of the Hamiltonian which shifts the momentum operator like
\begin{align}
 -i \partial_j\rightarrow -i \partial_j+\phi_j
\label{eqn:tbcshift} 
\end{align}
Using $\mathcal H_i = \mathcal H_{\phi_i}= \left[\partial_{\phi_i},\mathcal H\right]$~the Hall conductivity can be expressed as the sensitivity of the ground state wave function to TBC.
\begin{align*}
 \sigma_{xy}=\frac{2}{A}\textrm{Im}\langle\partial_{\phi_x} \Psi_0\lvert\partial_{\phi_y}\Psi_0\rangle,
\end{align*}
where $\lvert\Psi_0\rangle$~is the many body ground state after the mentioned gauge transformation which depends on the twisting angles. Defining $\theta=L_x \phi_x,~\varphi=L_y \phi_y$, 
\begin{align}
 \sigma_{xy}=2\textrm{Im}\langle\partial_\theta \Psi_0\lvert\partial_\varphi \Psi_0\rangle= i\mathcal  F_{\theta \varphi}.
 \label{eqn:halltbc}
\end{align}
The main merit of Ref. \onlinecite{Niu1985} is to show that this expression actually does not depend on the value of $(\theta,\varphi)$~as long as the single particle Green's function of the system is exponentially decaying in real space. This condition is met if the Fermi energy lies in a mobility gap. Hence, a trivial integration can be introduced as follows:
\begin{align}
\sigma_{xy}=\frac{i}{4\pi^2}\int_0^{2\pi}\textrm{d}\theta \int_0^{2\pi}\textrm{d}\varphi \mathcal F_{\theta\varphi}= G_0 \int_{T^2}\frac{i\mathcal F}{2\pi}=G_0 \mathcal C_1,
\label{eqn:niutbcchern}
\end{align}
where $G_0=\frac{e^2}{h}=\frac{1}{2\pi}$~and the integer $\mathcal C_1$~is by definition the first Chern number of the ground state line bundle over the torus of twisting angles. This construction makes the topological quantization of the Hall conductivity manifest.\\

Eq. (\ref{eqn:tbcshift}) shows the close relation between momentum and twisting angles. One is thus tempted to just replace the BZ of each periodic system by the torus of twisting angles for the corresponding disordered system which is topologically equivalent to a fictitious periodic system with the physical system as single lattice site \cite{FuInversion2007,ProdanTBC}. We will proceed along these lines below but would like to comment briefly on the special role played by the quantum Hall phase first. Eq. (\ref{eqn:halltbc}) represents the physical observable $\sigma_{xy}$~in terms of the twisting angles. Niu {\textit et al.} argued rigorously \cite{Niu1985} that  $\sigma_{xy}$~of a bulk insulating system can actually not depend on the value of these twisting angles which allows them to express $\sigma_{xy}$~as a manifestly quantized topological invariant in Eq. (\ref{eqn:niutbcchern}). For a generic TSM, the topological invariant of the clean system does in general not represent a physical observable. 
Furthermore, the 
integration over the 
twisting angles will not be trivial, i.e., the function to be integrated will actually depend on the twisting angles. Employing the picture of a periodic system with the physical system as a single site is problematic inasmuch as the bulk boundary correspondence at the ``boundary'' of a single site is hard to define mathematically rigorously. In the quantum Hall regime for example, it is well known that in disordered systems a complicated landscape of localized states and current carrying regions produces the unchanged topologically quantized Hall conductivity \cite{Halperin1982}. However, the edge states of the disordered quantum Hall state are in general not strictly localized at the boundary.\\

Replacing the BZ of a translation-invariant system by the torus of twisting angles in the disordered case yields a well defined topological invariant which adiabatically connects to the topological invariant of the clean system where the relation
\begin{align}
\partial_{k^j}=\partial_{\phi_j}
\end{align}
follows from Eq. (\ref{eqn:tbcshift}). This is because from a purely mathematical perspective it cannot matter which torus we consider as the base space of our system. In this sense, the framework of TBC is as good as it gets concerning the definition of topological invariants for disordered systems. When calculating the topological invariant of a symmetry protected TSM through dimensional extension (see Section \ref{sec:dimhier}), a hybrid approach between TBC in the physical dimensions and periodic boundaries in the extra dimension can be employed to define a topological invariant for the disordered system which can be calculated more efficiently \cite{z2maj}. The fact that in some symmetry protected topological phases the topological invariants are not directly representing physical observables is a not a problem of the approach of TBC but is a remarkable difference between these TSM and the quantum Hall state at a more fundamental level. Recently, an S-matrix approach to calculating topological invariants of non-interacting disordered TSM has been reported \cite{FulgaSmatrix} 

\subsection{Taking into account interactions}
\label{sec:int}
Up to now, our discussion has only been concerned with non-interacting systems. As a matter of fact, the entire classification scheme discussed in Sec. \ref{sec:symmclass} massively relies on the prerequisite that the Hamiltonian is a quadratic form in the field operators. The violation of this classification scheme for systems with two particle interactions has been explicitly demonstrated in Ref. \onlinecite{KitaevIntClass}.\\

As we are not able to give a general classification of TSM for interacting systems, we search for an adiabatic continuation of the non-interacting topological invariants to interacting systems. This procedure does from its outset impose certain adiabaticity constraints on the interactions that can be taken into account. The topological invariants for non-interacting systems are defined in terms of the projection $P$~on the occupied single particle states defining the ground state of the system.
The main assumption is thus that the gapped ground state of the non-interacting system is adiabatically connected to the gapped ground state of the interacting system. A counter-example of this phenomenology are fractional quantum Hall states \cite{StormerFQH,LaughlinState}, where a gap due to non-adiabatic interactions emerges in a system which is gapless without interactions. However, it is clear that the phase space for low energy interactions will be much larger in a gapless than in a gapped non-interacting system. We thus generically expect the classification scheme at hand to be robust against moderate interactions. However, beyond mean field interactions, the Hamiltonian cannot be expressed as an effective single particle operator.
Hence, we need to find a formulation of the topological invariants that adiabatically connects to the non-interacting language and is well defined for general gapped interacting systems. The key to achieving this goal is to look at the single particle Green's function $G$~instead of the Hamiltonian. This approach has been pioneered in the field of TSM by Qi, Hughes, and Zhang \cite{QiTFT}~who formulated a topological field theory for TSM in the CAZ class AII.\\
\subsubsection*{Chern numbers and Green's function winding numbers}
The role model for this construction is again the Hall conductivity $\sigma_{xy}$~of a gapped 2D system. In Ref. \onlinecite{Redlich1984}, $\sigma_{xy}$~has been expressed in terms of $G$~by perturbative expansion of the effective action of a gauge field $A$~that is coupled to the gapped fermionic system in the framework of quantum electrodynamics in (2+1)D. The leading contribution stemming from a vacuum polarization diagram yields the Chern Simons action
\begin{align*}
S_{\text{CS}}=\frac{\sigma_{xy}}{2}\int \text{d}^2x\text{d}t~\epsilon^{\mu\nu\sigma}A_\mu\partial_\nu A_\sigma=\frac{\sigma_{xy}}{2}\int A\wedge dA.
\end{align*}
The prefactor $\sigma_{xy}$~in units of the quantum of conductance assumes the form
\begin{align}
\sigma_{xy}= \frac{1}{24\pi^2}\int \text{d}^2k \text{d}\omega \text{Tr}\left[(GdG^{-1})^3\right],
\label{eqn:redlich}
\end{align}
where $d$~now denotes the exterior derivative in combined frequency-momentum space and $G$~is the time ordered Green's function, or, equivalently as far as the calculation of topological invariants is concerned, the continuous imaginary frequency Green's function as used in zero temperature perturbation theory. Eq. (\ref{eqn:redlich}) has first been identified as a topological invariant and been proven in a non-relativistic condensed matter context in Refs. \onlinecite{So1985,Ishikawa1986,Ishikawa1987}. An analogous expression has been derived by Volovik using a semi-classical gradient expansion \cite{VolovikQH3HE}. The similarity between Eq. (\ref{eqn:winddens}) and the integrand of Eq. (\ref{eqn:redlich}) is striking. Obviously, Eq. (\ref{eqn:redlich}) represents $\sigma_{xy}$~as a winding number in 3D frequency-momentum space. If this construction makes sense, we should by integration of Eq. (\ref{eqn:redlich}) over $\omega$~recover the representation of $\sigma_{xy}$~as the first Chern number in the 2D BZ for the special case of the non-interacting Green's function $G_0(\omega,k)=(i\omega -h(k))^{-1}$~. This straightforward calculation relies on the residue theorem and has been explicitly 
presented in Ref. \onlinecite{QiTFT}. Its result can be readily generalized to higher even spatial dimensions and higher Chern numbers, respectively. In Ref. \onlinecite{Golterman1993}, a perturbative expansion similar to Ref. \onlinecite{Redlich1984} has been presented for fermions coupled to a gauge field in arbitrary even spatial dimension $2n$. The resulting analogue of the Hall conductivity, i.e., the prefactor of the Chern Simons form in $(2n+1)$D (see Eq. (\ref{eqn:csdef})) can be expressed as \cite{Golterman1993,Volovik,QiTFT}
\begin{align}
\label{eqn:volovikgen}
&N_{2r+1}\left[G\right]=\mathcal N(r)\int_{\text{BZ}\times\mathbb R_\omega}\text{Tr}\left[\left(GdG^{-1}\right)^{2r+1}\right],\\
&\mathcal N(r)=\frac{-r!}{(2r+1)!(2\pi i)^{r+1}}.\nonumber
\end{align}
Performing again the integration over the frequency analytically for the noninteracting Green's function $G_0$~yields
\begin{align}
N_{2r+1}\left[G_0\right]=\mathcal C_r.
\label{eqn:volovikchernid}
\end{align}
Eq. (\ref{eqn:volovikchernid}) makes manifest that $N_{2r+1}\left[G\right]$, which can be formulated for an interacting system, reproduces the non-interacting classification for the free Green's function $G_0$~of the non-interacting system. The topological invariance of $N_{2r+1}\left[G\right]$~is clear by analogy with Eq. (\ref{eqn:windingdef}): Whereas the winding number $\nu_{2j-1}$~measures the homotopy of the chiral map $k\mapsto q(k)\in U(n)$~which, properly normalized, yields an integer due to $\pi_{2j-1}\left(U(n)\right)=\mathbb Z,~n>j$, Eq.  (\ref{eqn:volovikgen}) measures the homotopy of $G\in GL(n+m,\mathbb C)$~in the $(2r+1)$D frequency-momentum space which is also integer due to $\pi_{2r+1}\left(GL(n+m,\mathbb C)\right)=\mathbb Z,~n+m>r$.\\

The dimensional hierarchy for symmetry protected descendants of a parent state which is characterized by a Chern number (see Section \ref{sec:dimhier}) can be constructed in a completely analogous way for the interacting generalization $N_{2r+1}$~of the Chern number $\mathcal C_r$~\cite{TopologicalOrderParameter}. The resulting topological invariants for the descendant states have been coined topological order parameters in Ref. \onlinecite{TopologicalOrderParameter}. Disorder can again be accounted for by imposing TBC and replacing the $k$-space of the system by the torus of twisting angles (see Section \ref{sec:tbc}). Our discussion is limited to insulating systems. A detailed complementary analysis of the topological properties of different quantum vacua can be found in Ref. \onlinecite{VolovikReview}.
\subsubsection*{Interacting chiral systems}
The integer invariant of chiral unitary systems (class AIII) in odd spatial dimension $2r-1$~is not a Chern number but a winding number (see Section \ref{sec:complexclasses}). For all these systems and dimensional hierarchies with a chiral parent state, i.e., all chiral TSM (see Tab. \ref{tab:hierarchy}), a similar interacting extension of the definition of the invariants in terms of $G(i\omega,k)$~has been reported in Ref. \onlinecite{Gurarie2011}:
\begin{align}
I_{2r}\left[G\right]=n(r)\int_{\text{BZ}\times\mathbb R_\omega}\text{Tr}\left[Q\left(dQ\right)^{2r}\right],
\label{eqn:chiralG}
\end{align}
where $n(r)$~is a normalization constant, and $Q(i\omega,k)=G^{-1}(i\omega)U_{\textrm{CH}}G(i\omega,k)$, with the unitary representation matrix $U_{\textrm{CH}}$~of the chiral symmetry operation. In the non-interacting limit, $I_{2r}$~reduces to $\nu_{2r-1}$~as defined in Eq. (\ref{eqn:windingdef}) \cite{Gurarie2011}.

\subsubsection*{Fluctuation driven topological transitions}
Thus far, we have shown that for a non-interacting system the integration over $\omega$~reproduces the band structure classification scheme formulated in terms of the adiabatic curvature. However, the additional frequency dependence of the single particle Green's function can cause phenomena without non-interacting counterpart. To see this, we represent the single particle Green's function of an interacting system as
\begin{align*}
G(\omega,k)=\left(i\omega-h(k)-\Sigma(\omega,k)\right)^{-1},
\end{align*}
where $\Sigma$~is the self-energy of the interacting system. In Ref. \onlinecite{Gurarie2011}, it has been pointed out, that the value of $N_{2r+1}\left[G\right]$~cannot only change due to gap closings in the energy spectrum as in the case of the Chern number $\mathcal C_r$. This is due to the possibility of poles in the $\omega$-dependence of the self-energy which give rise to zeros of the Green's function, whereas gap closings correspond to poles of $G$. From the analytical form of $N_1$~it is immediately clear that both poles and zeros of $G$~can change the value of $N_1$. More generally, the $G\leftrightarrow G^{-1}$~symmetry of $N_{2r+1}$~makes clear that poles of $G$~can be seen as zeros of $G^{-1}$~and vice versa on an equal footing. In Ref. \onlinecite{FDWN2011}, it has been demonstrated that the $\omega$-dependence of $\Sigma$~can change a non-trivial winding number into a trivial one. The emergence of a topologically nontrivial phase due to dynamical fluctuations which has no non-interacting counterpart has 
been presented in Refs. \onlinecite{localself,fluchund}.

\subsubsection*{Chern numbers of effective single particle Hamiltonians}

Due to its additional $\omega$-integration, the practical calculation of $N_{2r+1}$~can be numerically very challenging once the single particle Green's function of the interacting system has been calculated. A major breakthrough along these lines has been the observation that an effective single particle Hamiltonian defined in terms of the inverse Green's function at zero frequency can be defined to effectively reduce the topological classification to the non-interacting case. This possibility has first been mentioned in Refs. \onlinecite{VolovikStandard,VolovikSoftTop} and been generally proven in Ref. \onlinecite{WangGeneralTOP}. The authors of Ref. \onlinecite{WangGeneralTOP} show, using the spectral representation of the Green's function, that one can always get rid of the $\omega$-dependence of $G$. We only review the physical results of this analysis. The both accessible and explicit proof can be found in Ref. \onlinecite{WangGeneralTOP}. The physical conclusion is as elegant as simple: Instead of calculating $N_{2r+1}$~we can just calculate the Chern number $\mathcal C_r$~associated with the fictitious Hamiltonian
\begin{align}
\tilde h(k)=-G^{-1}(0,k),
\label{eqn:Wangham}
\end{align}
the occupied states of which are just its eigenstates with negative eigenvalues which have been dubbed $R$-zeros \cite{WangGeneralTOP} since they are positive energy eigenstates of $G^{-1}(0,k)$. Obviously, $-G_0^{-1}(0,k)=h(k)$~for the non-interacting Green's function. Hence, $\tilde h(k)$, which has recently been coined topological Hamiltonian \cite{topham}, adiabatically connects to $h(k)$~in the non-interacting limit. Note that the possibility of eliminating the $\omega$-dependence is not in contradiction to the relevance of this $\omega$-dependence for the topology of the interacting system. All it shows is that the relevant changes due to a different pole structure of $G$~as a function of $\omega$~can be inferred from its value at $\omega=0$.

\subsubsection*{Topological Hamiltonian for chiral interacting systems}
In principle, the construction of the topological Hamiltonian $\tilde h(k)$~can be readily generalized to chiral interacting systems as has been mentioned in Ref. \onlinecite{WangTSC}. To see this, we note that the crucial argument for the construction of the topological Hamiltonian brought forward in Ref. \onlinecite{WangGeneralTOP} is the following: The continuous interpolation
\begin{align*}
G(i\omega,k,\lambda) = (1-\lambda)G(i\omega,k)+\lambda \left[i\omega +G^{-1}(0,k)\right]^{-1}
\end{align*}
does not contain any singularities or gap closings. Thus, as long as the calculation of a topological invariant in terms of $G(i\omega,k)$~is concerned, we can also use $\tilde G(i\omega,k)= G(i\omega,k,\lambda=1)=\left[i\omega +G^{-1}(0,k)\right]^{-1}$. Obviously, $\tilde G(i\omega,k)=\left[i\omega -\tilde h(k)\right]^{-1}$~is the Green's function of a fictitious non-interacting system which is governed by the topological Hamiltonian $\tilde h(k)$. The mere existence of the topological invariant for chiral systems in terms of $G(i\omega,k)$~as presented in Ref. \onlinecite{Gurarie2011} (see also Eq. (\ref{eqn:chiralG})) hence suffices to argue that one can equally well investigate the topology of $\tilde h(k)$~and its symmetry protected descendants (see Section \ref{sec:dimhier}) instead of directly evaluating Eq. (\ref{eqn:chiralG}).
Since the single particle Green's function inherits the fundamental symmetries from the Hamiltonian \cite{Gurarie2011}, $\tilde h(k)$~will also obey these symmetries. In particular, for an interacting system with chiral symmetry, the topological Hamiltonian can be brought into the flat band off-diagonal form
\begin{align}
\tilde h(k)\simeq\begin{pmatrix}
             0&{\tilde q(k)}\\
             {\tilde q^\dag(k)}&0
            \end{pmatrix},
\label{eqn:fictitiouschiral}
\end{align}
where $\tilde q(k)\in U(n)$~for the topologically equivalent flat-band system. This construction generically defines a topological invariant for the chiral interacting system which adiabatically concurs with the non-interacting system: The winding number $\tilde \nu$~associated with the fictitious Hamiltonian $\tilde h(k)$. A similar construction for a chiral 1D system has been presented in Ref. \onlinecite{ZGurarie}.\\

\subsubsection*{Discussion of the topological Hamiltonian and practical consequences}
As already mentioned, the above construction cannot be valid for arbitrary interacting systems. In particular in 1D, the breakdown of the $\mathbb Z$~classification in the presence of general interactions has been investigated in Ref. \onlinecite{KitaevIntClass}. However, this problem does not pertain to the concept of the topological Hamiltonian itself but rather reflects the limited validity of the adiabatic continuation of the non-interacting invariants in terms of the single particle Green's function, i.e., the limited validity of Eq. (\ref{eqn:volovikgen}) and Eq. (\ref{eqn:chiralG}). In the validity regime of these equations, one can equivalently use the topological Hamiltonian $\tilde h(k)=-G^{-1}(0,k)$~to classify an interacting system in any symmetry class. This is of enormous practical usefulness for at least two reasons. First, we get rid of the $\omega$-integration appearing in Eq. (\ref{eqn:volovikgen}) and Eq. (\ref{eqn:chiralG}) which is cumbersome to evaluate. Second, the 
method of dimensional 
extension, though generally valid, is not always the most convenient way to calculate the topological invariant of a symmetry protected descendant state. Provided with the formal equivalence between the non-interacting classification problem of the topological Hamiltonian and the Green's function topology, we can directly apply all simplified schemes that have been introduced to directly calculate non-interacting invariants of symmetry protected states (see, e.g., Refs. \onlinecite{Kitaev2001,FuPump,FuInversion2007,Prodanz2,WangInversion}) to the topological Hamiltonian. The framework of dimensional extension and Eq. (\ref{eqn:volovikgen}) or Eq. (\ref{eqn:chiralG}) for the parent state are, with the benefit of hindsight, only needed to justify the validity of the topological Hamiltonian.\\

Before closing the section, we would like to discuss the role of the bulk boundary correspondence, in the presence of interactions. In general, interactions can spontaneously break the protecting symmetry of a symmetry protected TSM locally at the boundary thus gapping out the characteristic metallic surface states. Importantly, this spontaneous symmetry breaking at the gapless surface will typically happen at a lower critical interaction strength than in the gapped bulk. This is because the gapless surface modes offer more phase space for interactions.  A generally valid bulk boundary correspondence is hence absent in the interacting case. Within the validity regime of Eq. (\ref{eqn:volovikgen}) and Eq. (\ref{eqn:chiralG}) for chiral TSM, respectively, an interacting analogue of the bulk boundary correspondence has been reported in Refs. \onlinecite{Gurarie2011,EssinGurarie}. The main difference to the non-
interacting case is that boundary zero-modes, which represent poles of the Green's function can be canceled by zeros of the Green's function as far as the calculation of topological invariants is concerned. Note that the Green's function of a non-interacting system does not have zeros.
\section{Limitations of the framework of TSM}
Finally, we would like to point out some limitations of the concept of TSM. The two main aspects that one could see critical in the field of TSM are outlined in the following.\\

First, whereas the topologically quantized Hall conductivity in the integer quantum Hall state, the historical role model of all TSM, is a physical observable, the topological invariants of symmetry protected TSM like the QSH state are not directly physically observable without additional unitary symmetries. The quantum Hall effect can be understood in terms of the spectral flow associated with the threading of a flux tube \cite{AvronIndex}. Along similar lines, the QSH effect can be understood in terms of a spin charge separation associated with the threading of a spin flux \cite{QiSpinChargeSep}. However, this spin flux, as opposed to an ordinary magnetic flux tube, is not immediately experimentally accessible and the general observable consequences of the QSH state have been shown to be much more subtle \cite{QSHSignatures}. For several TSM, the directly measurable consequences of the respective topological invariants are still under debate or unknown.\\

Second, the entire construction and classification of TSM is based on single particle Hamiltonians. In Section \ref{sec:int}, we discussed how adiabatic interactions can be taken into account and argued that interactions of moderate strength are not likely to destroy the phenomenology of TSM. In order to position the field of TSM in a broader context, we would like to point out that there are also phenomena of topological origin which emerge only due to the presence of interactions.
The historically first phenomenon is the $\frac{1}{\nu}$~FQH effect \cite{StormerFQH,LaughlinState} which cannot be adiabatically connected to an insulating non-interacting state. The non-interacting state is in this case a partially filled Landau level which provides an enormous density of states at the Fermi energy. In a system with periodic boundaries, the $\frac{1}{\nu}$~FQH has a characteristic $\nu$-fold ground state degeneracy. Interestingly, this simplest FQH state can still be analyzed in the framework of TBC \cite{Niu1985}. For more general FQH systems the concept of topological order has been introduced by Wen \cite{WenTO}. A crucial notion in this framework is the quantum dimension of the topologically ordered system which can be viewed as the ground state degeneracy of the system on a torus, i.e., with periodic boundary conditions. Most TSM discussed in this article have quantum dimension one, like a trivial insulator. From the vantage point of topological order, these states are thus trivial. Recently, the concepts of TSM and FQH physics have been combined to the definition of the fractional Chern insulator \cite{wenfci,sunfci,neupertfci} and the fractional topological insulator \cite{bernevig2006a,LevinFractionalTI,ZhangFractionalTI,maissamfracti} the first lattice realization of which has been reported in Ref. \onlinecite{Mudrylattice}. These states are translation-invariant realizations of the FQH effect and its time reversal symmetry protected analogues, respectively. A general hierarchy of fractional topological insulators has been reported in Ref. \onlinecite{MudryHierarchy}.

\section{Outlook}
From a conceptual point of view, the entire zoo of topological states of matter can be seen as conclusively understood in the framework outlined in this review. However, there are at least two general routes to be considerably further explored by future research. First, the precise experimental implications of many topological states of matter have not been fully unraveled yet. Whereas in the quantum Hall state, the topological invariant directly represents a physical observable, namely the Hall conductivity of the sample, the observability of the topological invariants of several topological states of matter is unknown or still under active debate. This issue is from our point of view closely related to the rather limited number of promising proposals for concrete technological applications based on these novel states of matter. Obviously, successful research in this direction will be of decisive importance for the long term future of 
the entire field of topological states of matter. Second, the influence of interactions and open quantum system effects on topological states of matter is by no means conclusively understood, let alone an exhaustive topological classification of interacting or dissipative systems. As a first step along these lines, a purely dissipation driven topological state has been reported in Ref. \onlinecite{DiehlNaturePhysics,DiehlZoller}.\\

\section*{acknowledgement}
We acknowledge interesting discussions with Hans Hansson, Patrik Recher, Dietrich Rothe, Ronny Thomale, Grigori Volovik, and Shou-Cheng Zhang as well as financial support from the DFG-JST research unit ``Topotronics'' (BT and JCB) and from the Swedish Science Research Council (JCB).

\end{document}